\pdfoutput=1  % Instructs arXiv to run this with pdflatex rather than latex.

\documentclass[aps,twocolumn,groupedaddress,floatfix,preprintnumbers,nofootinbib,eqsecnum]{revtex4}

\usepackage{amsmath, float, graphicx,placeins,amssymb,multirow,pgfplots,pgfplotstable}
\usepackage{tikz,hyperref}
\usetikzlibrary{shapes.geometric,arrows}

\begin{document}
\title{
Boxes, Boosts, and Energy Duality: \\
Understanding the Galactic-Center Gamma-Ray Excess \\
through Dynamical Dark Matter
}

\author{Kimberly K.\ Boddy$^{1}$\footnote{E-mail address:  {\tt kboddy@hawaii.edu}},
  Keith R.\ Dienes$^{2,3}$\footnote{E-mail address:  {\tt dienes@email.arizona.edu}},
  Doojin Kim$^{4,5}$\footnote{E-mail address:  {\tt doojin.kim@cern.ch}},\\
  Jason Kumar$^{1}$\footnote{E-mail address:  {\tt jkumar@hawaii.edu}},
  Jong-Chul Park$^{6}$\footnote{E-mail address:  {\tt jcpark@cnu.ac.kr}},
  Brooks Thomas$^{7}$\footnote{E-mail address:  {\tt thomasbd@lafayette.edu}}\\ ~}
\affiliation{
  $^1\,$Department of Physics \& Astronomy, University of Hawaii, Honolulu, HI 96822  USA\\
  $^2\,$Department of Physics, University of Arizona, Tucson, AZ  85721  USA\\
  $^3\,$Department of Physics, University of Maryland, College Park, MD  20742  USA\\
  $^4\,$Department of Physics, University of Florida, Gainesville, FL 32611  USA\\
  $^5\,$Theory Division, CERN, CH-1211 Geneva 23, Switzerland\\
  $^6\,$Department of Physics, Chungnam National University, Daejeon 34134  Korea\\
  $^7\,$Department of Physics, Lafayette College, Easton, PA  18042  USA}

\begin{abstract}
Many models currently exist 
which attempt to interpret
the excess of gamma rays emanating from the Galactic Center in terms of annihilating or decaying dark matter.
These models typically exhibit a variety of complicated cascade mechanisms for photon production,
leading to a non-trivial kinematics which obscures the physics of the underlying dark sector.
In this paper, by contrast, we observe that the spectrum of the gamma-ray excess may actually
exhibit an intriguing ``energy-duality'' invariance under $E_\gamma \rightarrow E_\ast^2/E_\gamma$ for some $E_\ast$.
As we shall discuss, such an energy duality points back to a remarkably simple alternative kinematics
which in turn is realized naturally within the Dynamical Dark Matter framework.
Observation of this energy duality could therefore provide considerable information
about the properties of the dark sector
from which the Galactic-Center gamma-ray excess might arise,
and highlights the importance of acquiring more complete data for the Galactic-Center excess in the energy range around 1~GeV.
\end{abstract}

\maketitle

%========================================================================
%          KEYSROKE-SAVING MACROS, nothing complicated
%========================================================================

\newcommand{\PRE}[1]{{#1}} % Use if preprint style
\newcommand{\ul}{\underline}
\newcommand{\del}{\partial}
\newcommand{\nbox}{{\,\lower0.9pt\vbox{\hrule \hbox{\vrule height 0.2 cm
\hskip 0.2 cm \vrule height 0.2 cm}\hrule}\,}}

\newcommand{\postscript}[2]{\setlength{\epsfxsize}{#2\hsize}
   \centerline{\epsfbox{#1}}}
\newcommand{\gweak}{g_{\text{weak}}}
\newcommand{\mweak}{m_{\text{weak}}}
\newcommand{\mplanck}{M_{\text{Pl}}}
\newcommand{\mstar}{M_{*}}
\newcommand{\sigmaan}{\sigma_{\text{an}}}
\newcommand{\sigmatot}{\sigma_{\text{tot}}}
\newcommand{\sigmaSI}{\sigma_{\rm SI}}
\newcommand{\sigmaSD}{\sigma_{\rm SD}}
\newcommand{\OmegaM}{\Omega_{\text{M}}}
\newcommand{\OmegaDM}{\Omega_{\text{DM}}}
\newcommand{\ipb}{\text{pb}^{-1}}
\newcommand{\ifb}{\text{fb}^{-1}}
\newcommand{\iab}{\text{ab}^{-1}}
\newcommand{\ev}{\text{eV}}
\newcommand{\kev}{\text{keV}}
\newcommand{\mev}{\text{MeV}}
\newcommand{\gev}{\text{GeV}}
\newcommand{\tev}{\text{TeV}}
\newcommand{\pb}{\text{pb}}
\newcommand{\mb}{\text{mb}}
\newcommand{\cm}{\text{cm}}
\newcommand{\m}{\text{m}}
\newcommand{\km}{\text{km}}
\newcommand{\kg}{\text{kg}}
\newcommand{\g}{\text{g}}
\newcommand{\s}{\text{s}}
\newcommand{\yr}{\text{yr}}
\newcommand{\Mpc}{\text{Mpc}}
\newcommand{\etal}{{\em et al.}}
\newcommand{\eg}{{\em e.g.}}
\newcommand{\ie}{{\em i.e.}}
\newcommand{\ibid}{{\em ibid.}}
\newcommand{\Eqref}[1]{Equation~(\ref{#1})}
\newcommand{\secref}[1]{Sec.~\ref{sec:#1}}
\newcommand{\secsref}[2]{Secs.~\ref{sec:#1} and \ref{sec:#2}}
\newcommand{\Secref}[1]{Section~\ref{sec:#1}}
\newcommand{\appref}[1]{App.~\ref{sec:#1}}
\newcommand{\figref}[1]{Fig.~\ref{fig:#1}}
\newcommand{\figsref}[2]{Figs.~\ref{fig:#1} and \ref{fig:#2}}
\newcommand{\Figref}[1]{Figure~\ref{fig:#1}}
\newcommand{\tableref}[1]{Table~\ref{table:#1}}
\newcommand{\tablesref}[2]{Tables~\ref{table:#1} and \ref{table:#2}}
\newcommand{\Dsle}[1]{\slash\hskip -0.28 cm #1}
\newcommand{\met}{{\Dsle E_T}}
\newcommand{\mpt}{\not{\! p_T}}
\newcommand{\Dslp}[1]{\slash\hskip -0.23 cm #1}
\newcommand{\Dsl}[1]{\slash\hskip -0.20 cm #1}

\newcommand{\mB}{m_{B^1}}
\newcommand{\mq}{m_{q^1}}
\newcommand{\mf}{m_{f^1}}
\newcommand{\mKK}{m_{KK}}
\newcommand{\WIMP}{\text{WIMP}}
\newcommand{\SWIMP}{\text{SWIMP}}
\newcommand{\NLSP}{\text{NLSP}}
\newcommand{\LSP}{\text{LSP}}
\newcommand{\mWIMP}{m_{\WIMP}}
\newcommand{\mSWIMP}{m_{\SWIMP}}
\newcommand{\mNLSP}{m_{\NLSP}}
\newcommand{\mchi}{m_{\chi}}
\newcommand{\mgravitino}{m_{\gravitino}}
\newcommand{\mmed}{M_{\text{med}}}
\newcommand{\gravitino}{\tilde{G}}
\newcommand{\Bino}{\tilde{B}}
\newcommand{\photino}{\tilde{\gamma}}
\newcommand{\stau}{\tilde{\tau}}
\newcommand{\slepton}{\tilde{l}}
\newcommand{\snu}{\tilde{\nu}}
\newcommand{\squark}{\tilde{q}}
\newcommand{\mgaugino}{M_{1/2}}
\newcommand{\epsEM}{\varepsilon_{\text{EM}}}
\newcommand{\mmess}{M_{\text{mess}}}
\newcommand{\lmess}{\Lambda}
\newcommand{\nmess}{N_{\text{m}}}
\newcommand{\signmu}{\text{sign}(\mu)}
\newcommand{\Omegachi}{\Omega_{\chi}}
\newcommand{\lambdafs}{\lambda_{\text{FS}}}
\newcommand{\be}{\begin{equation}}
\newcommand{\ee}{\end{equation}}
\newcommand{\bea}{\begin{eqnarray}}
\newcommand{\eea}{\end{eqnarray}}
\newcommand{\beq}{\begin{equation}}
\newcommand{\eeq}{\end{equation}}
\newcommand{\beqn}{\begin{eqnarray}}
\newcommand{\eeqn}{\end{eqnarray}}
\newcommand{\baln}{\begin{align}}
\newcommand{\ealn}{\end{align}}
\newcommand{\lsim}{\lower.7ex\hbox{$\;\stackrel{\textstyle<}{\sim}\;$}}
\newcommand{\gsim}{\lower.7ex\hbox{$\;\stackrel{\textstyle>}{\sim}\;$}}

\newcommand{\ssection}[1]{{\em #1.\ }}
\newcommand{\rem}[1]{\textbf{#1}}

\def\ie{{\it i.e.}\/}
\def\eg{{\it e.g.}\/}
\def\etc{{\it etc}.\/}
\def\calN{{\cal N}}

\def\mptwo{{m_{\pi^0}^2}}
\def\mp{{m_{\pi^0}}}
\def\sqtsn{\sqrt{s_n}}
\def\sqtsn{\sqrt{s_n}}
\def\sqtsn{\sqrt{s_n}}
\def\sqts0{\sqrt{s_0}}
\def\Dsqts{\Delta(\sqrt{s})}
\def\Omegatot{\Omega_{\mathrm{tot}}}

%%%%%%%%%%%%%%%%%%%%%%%%%%%%%%%%%%%%%%%%%%%%%%%%%%%%%%%%%%%%%%%%%%%%%%%%%%%%%%%

\section{Introduction}

%%%%%%%%%%%%%%%%%%%%%%%%%%%%%%%%%%%%%%%%%%%%%%%%%%%%%%%%%%%%%%%%%%%%%%%%%%%%%%%

A robust excess in the flux of gamma-ray photons emanating from the Galactic Center 
(GC) with energies of $\mathcal{O}(\mathrm{GeV})$ has been observed in Fermi Large Area 
Telescope (Fermi-LAT) data.  This excess was first noted in Ref.~\cite{Goodenough:2009gk} 
and corroborated by a number of subsequent, independent analyses~\cite{Hooper:2010mq,
Hooper:2011ti,Abazajian:2012pn,Hooper:2013rwa,Gordon:2013vta,Huang:2013pda,
Abazajian:2014fta,Daylan:2014rsa,Lacroix:2014eea,
YizhongGC,
Calore:2014xka,Calore:2014nla}, including 
a dedicated study by the Fermi-LAT collaboration itself~\cite{TheFermi-LAT:2015kwa}.
This excess consists not of a spectral line, but rather of a continuum bump which extends 
over a range of photon energies $0.3\mbox{~GeV}\lesssim E_\gamma \lesssim 50\mbox{~GeV}$ 
and peaks at approximately $E_\gamma \sim 1$~GeV.

A variety of possible explanations have been advanced as to the origin of this gamma-ray 
excess.  Possible astrophysical explanations include emission from a population of 
millisecond pulsars~\cite{Hooper:2010mq,Hooper:2011ti,Abazajian:2012pn,Gordon:2013vta,
Abazajian:2014fta,Abazajian:2010zy} and the decay of neutral pions produced by collisions 
of cosmic-ray particles with interstellar gas~\cite{Hooper:2010mq,Hooper:2011ti,
Abazajian:2012pn,Gordon:2013vta}.  However, the spectrum produced by millisecond pulsars 
is too soft in the sub-GeV region to explain the observed data~\cite{Hooper:2013nhl} and
millisecond pulsars born in globular clusters can account for only a few percent or less 
of the observed excess~\cite{Hooper:2016rap}.  In addition, the observed distributions of 
gas seem to yield a poor fit to the spatial distribution of the 
signal~\cite{Lacroix:2014eea,Linden:2012iv,Macias:2013vya}.  More recently, in 
Refs.~\cite{Bartels:2015aea,Lee:2015fea}, it has been asserted that the excess can be 
described by a set of unresolved point sources, and new methods to characterize these 
sources were devised.

An exciting alternative possibility is that the excess is the result of annihilating or
decaying dark-matter particles within the galactic halo.  Indeed, the spatial morphology 
of the excess is consistent with dark-matter annihilations from a spherically symmetric 
density profile, and the excess extends outward more than $10^\circ$ from its center at 
the dynamical center of the Milky Way~\cite{Daylan:2014rsa}.  As a result, a variety of 
models currently exist in the literature which posit a dark-matter origin for the 
continuum feature observed in the Fermi-LAT data.

In such models, a suitably broad spectrum of gamma rays is realized through a variety 
of complicated cascade mechanisms.  For example, such a gamma-ray spectrum can be 
generated via the subsequent showering and/or hadronization of Standard-Model (SM) 
particles initially produced directly from dark-matter annihilation.  The observed 
excess is well reproduced by a dark-matter (DM) particle with a mass $m_\chi \sim (30-50)$~GeV and an
annihilation cross-section 
$\langle\sigma v\rangle \approx (1-3)\times 10^{-26}~\hbox{cm}^3/\hbox{s}$ which 
annihilates primarily to $b\bar{b}$~\cite{Daylan:2014rsa,Calore:2014nla}.  Likewise, 
a dark-matter particle with a mass $m_\chi \sim 10$~GeV and an annihilation cross-section 
$\langle\sigma v\rangle \approx (0.5-2)\times 10^{-26} ~\hbox{cm}^3/\hbox{s}$ which 
annihilates primarily to $\ell^+\ell^-$~\cite{Lacroix:2014eea} also provides a good fit 
to the Fermi-LAT data, provided that secondary photons produced by inverse Compton 
scattering and bremsstrahlung processes involving both primary and secondary electrons 
are taken into account.  On the other hand, the recent AMS-02 data on the cosmic-ray 
antiproton flux~\cite{Aguilar:2015ooa, AMS02} has begun to exclude states in which a 
$q\bar{q}$ final state dominates~\cite{Giesen:2015ufa}.  Concrete models in which the 
dark-matter candidate annihilates primarily to $b\bar{b}$~\cite{Alvares:2012qv,
Okada:2013bna,Modak:2013jya,Alves:2014yha,Ipek:2014gua,Basak:2014sza} and to 
$\ell^+\ell^-$~\cite{Kyae:2013qna,Kim:2015fpa} have also been identified.
Indeed, additional studies on other final states~\cite{Calore:2014nla} and generic 
model constraints~\cite{Kong:2014haa} have established that there exist further
SM channels through which a dark-matter particle can annihilate or decay and reproduce 
the observed excess.  Cascades involving one or more exotic intermediary particles 
which eventually decay down to SM fermions which in turn subsequently shower or hadronize 
have also been considered~\cite{Boehm:2014bia,Ko:2014gha,Abdullah:2014lla,SheltonDMCascade,Elor:2015tva}.

While dark-matter models of this sort are capable of reproducing the GC excess, the 
showering and cascade dynamics on which these models rely in order to generate an 
acceptable gamma-ray spectrum have their disadvantages as well.  For example, their 
complicated dynamics obscures the relationship between the detailed shape of the 
gamma-ray spectrum and the properties of the underlying dark sector.

In this paper, by contrast, we propose a set of models in which the kinematics 
connecting the gamma-ray spectrum back to the dark sector is more straightforward.
As a result, we find that characteristic imprints in the shape of that spectrum 
can potentially provide direct information about the dark sector.  

We begin our study 
by identifying a characteristic feature of the GC gamma-ray excess which points back 
to a particularly simple photon-production kinematics.  In particular, we observe that 
the spectrum of this excess may potentially exhibit an intriguing ``energy duality'' 
under which the spectrum remains invariant under the transformation 
$E_\gamma\rightarrow E_\ast^2/E_\gamma$ for some self-dual energy $E_\ast$.  As we shall argue,
the presence of such an energy duality is indicative of a particularly simple kinematics 
in which the signal photons are produced directly via the two-body decays of an 
intermediary particle.  

Energy dualities of this sort have been exploited in other 
contexts involving similar decay kinematics, such as cosmic-ray pion 
decay~\cite{Stecker} and the decay of heavy (new) particles produced at 
colliders~\cite{KaustubhEnergyPeak}.  At present, due to uncertainties in the 
astrophysical modeling of the GC region and also due to a paucity of reliable 
information about the shape of the spectrum at photon energies 
$\mathcal{O}(10\mbox{~MeV}) \lesssim E_\gamma \lesssim \mathcal{O}(1\mbox{~GeV})$,
the information contained in the Fermi-LAT data alone is not sufficient to 
conclusively determine whether the spectrum of the GC excess in fact displays such 
an energy duality.  Nevertheless, as we shall discuss, if such a duality {\it were}\/ to 
be confirmed through future experiments, this result would immediately favor a particular class of
dark-matter models.  Moreover, these observations apply not only for the GC 
gamma-ray spectrum but also for the spectra from other sources, such as dwarf galaxies, 
for which backgrounds can be more reliably estimated.

While a spectrum with these duality properties can be realized in certain cascade-based 
models~\cite{DoojinAstroEnergyPeak}, we shall show that a self-dual gamma-ray spectrum
also has a natural interpretation within the Dynamical Dark Matter (DDM) 
framework~\cite{DDM1,DDM2}.  Indeed, as we shall show, there exists a simple class of 
DDM models which yield an energy-dual spectrum that provides an excellent fit to the 
Fermi-LAT data, with a self-dual energy $E_\ast \sim \mathcal{O}(1\mbox{~GeV})$.
These results further highlight the importance of acquiring more complete gamma-ray data 
in the energy range $10\mbox{~MeV} \lesssim E_\gamma \lesssim 1\mbox{~GeV}$.

This paper is organized as follows.  In Sect.~\ref{sec:GCExcess}, we examine the 
energy spectrum of the GC excess and discuss the extent to which it might potentially 
exhibit an energy-duality invariance under $E_\gamma\rightarrow E_\ast^2/E_\gamma$ with 
$E_\ast \sim \mathcal{O}(1\mbox{~GeV})$.  In Sect.~\ref{sec:Framework}, we then 
discuss how a gamma-ray spectrum with such an invariance can arise from dark-matter 
annihilation or decay.  In Sect.~\ref{sec:Models}, we introduce a series of simple DDM 
models which give rise to a gamma-ray spectrum with this invariance and demonstrate 
that such DDM models provide a successful fit to the Fermi-LAT data.
Our conclusions are then presented in Sect.~\ref{sec:conclusion}, where we 
also discuss the potential implications of energy duality for other astrophysical gamma-ray signals
which might be observed in the future.  Finally, an Appendix contains a derivation 
of certain results presented in the text.

%%%%%%%%%%%%%%%%%%%%%%%%%%%%%%%%%%%%%%%%%%%%%%%%%%%%%%%%%%%%%%%%%%%%%%%%%%%%%%%

\section{Energy Duality and the Galactic-Center Excess\label{sec:GCExcess}}

%%%%%%%%%%%%%%%%%%%%%%%%%%%%%%%%%%%%%%%%%%%%%%%%%%%%%%%%%%%%%%%%%%%%%%%%%%%%%%%

As discussed in the Introduction, there is evidence of an unexplained gamma-ray 
excess from the GC in the Fermi-LAT data near 1 GeV that may be due to DM 
annihilations or decays.  We consider the analysis in Ref.~\cite{Daylan:2014rsa}, 
in which the gamma-ray excess has been identified out to at least $10^\circ$ from 
the GC.  The authors of  Ref.~\cite{Daylan:2014rsa} fit the Fermi-LAT data to 
background templates consisting of the Galactic and extragalactic diffuse emission 
and the Fermi Bubbles.  They also include a potential signal template for dark-matter
annihilations.  This latter contribution may be written in terms of a differential-flux
contribution from dark-matter annihilation.  In a single-particle dark-matter scenario, 
this flux ${\cal F}$ may be written in the form
\begin{equation}
  {\cal F} \equiv \frac{d^2\Phi}{dE_\gamma\, d\Omega} = \frac{\mathcal{J}}{4\pi}
  \frac{\langle\sigma v\rangle}{4m^2} \frac{dN_\gamma}{dE_\gamma} \ ,
\label{fluxeq}
\end{equation}
where $m$ is the mass of the DM particle and $\langle\sigma v\rangle$ is its
velocity-averaged 
annihilation cross section.  Here $\mathcal{J} = \int d\ell \rho^2$, where $\rho$ is 
the dark-matter energy density and the integral is along the line of sight.  
Since the only non-trivial angular dependence for $\mathcal{F}$ arises from $\mathcal{J}$,
we may replace ${\cal J}$ in Eq.~(\ref{fluxeq}) 
by its angular average $\overline{\cal J}$ over
a relevant angle $\Delta \Omega$ on the sky, where
\beq
     \overline{\cal J} ~\equiv~ {1\over \Delta \Omega} \int_{\Delta \Omega} d\Omega \, {\cal J}~.
\eeq
We then find that
\beq
     {d \Phi\over dE_\gamma} ~=~ {J\over 4\pi} {\langle \sigma v \rangle \over 4 m^2} {d N_\gamma\over dE_\gamma}
\label{Jversion}
\eeq
where $J\equiv (\Delta \Omega) \overline{\cal J}$. 
In writing Eq.~(\ref{fluxeq}) we have 
assumed that the dark-matter particle is distinct from the antiparticle; if the particle 
and antiparticle are identical, the flux is rescaled by a factor of 2.  The energy 
density $\rho$ is assumed to follow a generalized Navarro-Frenk-White (NFW) halo 
profile~\cite{Navarro:1995iw, Navarro:1996gj}
\begin{equation}
  \rho(r) ~=~ \rho_0\frac{(r/r_s)^{-\gamma}}{(1+r/r_s)^{3-\gamma}}~,
  \label{eq:nfw}
\end{equation}
where $\rho_0 \simeq 0.4~\mathrm{GeV/cm^3}$ is the local DM density at $r\simeq8.5$~kpc 
and where $r_s=20$ kpc is the scale radius.  It is then found~\cite{Daylan:2014rsa}  that 
inclusion of this additional template with an inner-profile slope in the range 
$\gamma\approx 1.1$--$1.3$ significantly improves the overall fit, with the dark-matter
contribution taking the form of a continuum bump which peaks around $E_\gamma \sim 1$~GeV.

%=======================================================================================
\begin{figure}[t]
  \centering
  \includegraphics[width=8.7cm]{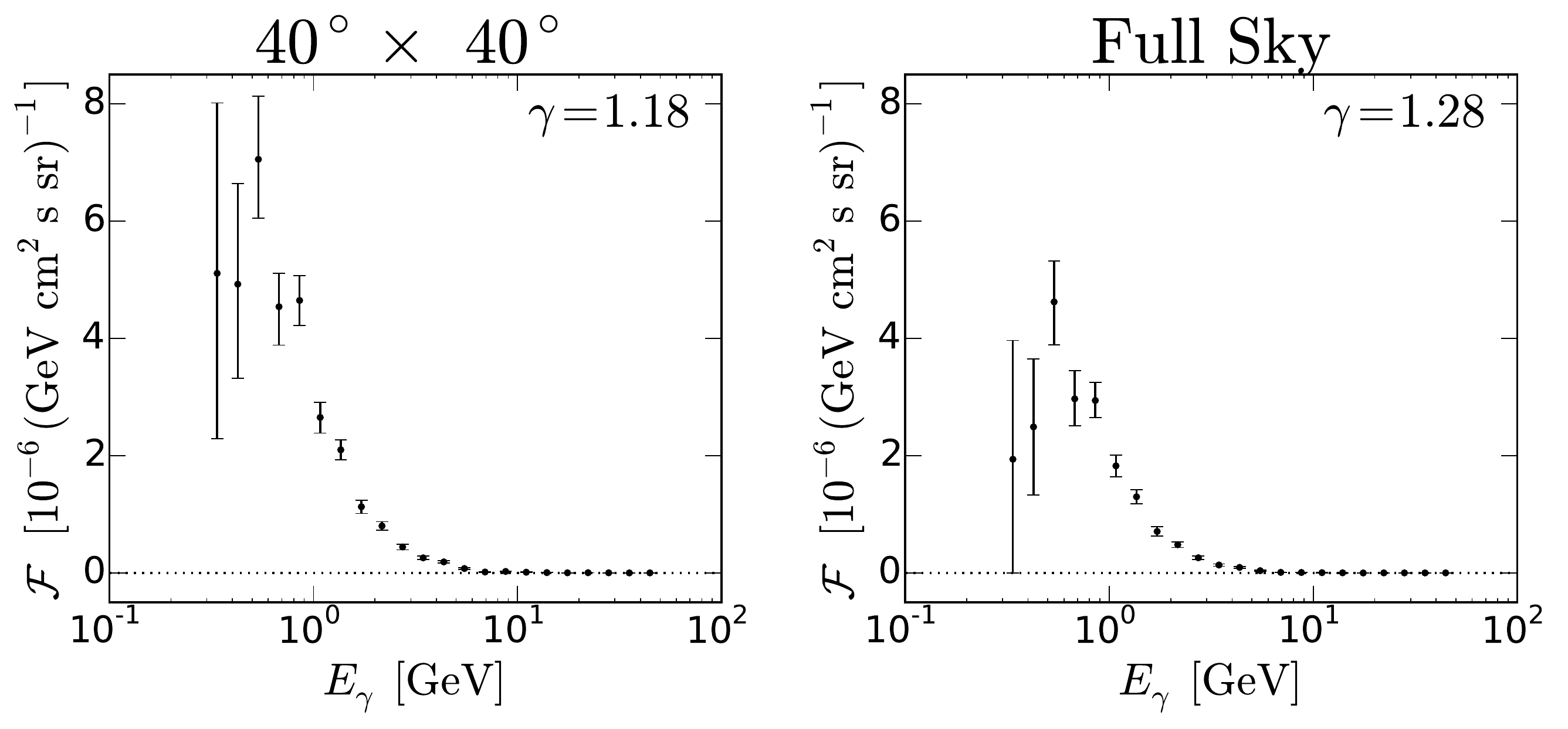}
  \caption{The observed GC flux excess ${\cal F}$.
    The data points and the associated error bars are extracted from 
    Ref.~\cite{Daylan:2014rsa}, while the values of $\gamma$ indicated in each panel 
    are the best-fit inner slopes of the generalized NFW profile in Eq.~\eqref{eq:nfw}.
    Note that these two panels correspond to different regions on the sky and thus
    correspond to different values of $\overline{\cal J}$.  }
  \label{fig:GC_flux}
\end{figure}
%=======================================================================================

This dark-matter excess is shown in Fig.~\ref{fig:GC_flux}, where we plot the 
residuals of the differential photon flux obtained from the analysis in 
Ref.~\cite{Daylan:2014rsa} of Fermi-LAT data from the GC.  In this analysis, each 
photon is placed into one of 22 energy bins, equally spaced on a logarithmic scale 
between 0.3 and 50~GeV.  We emphasize that the error bars in Fig.~\ref{fig:GC_flux} are 
statistical only, and that there are also large astrophysical uncertainties from 
background modeling which are not shown.  Note that there are two regions of 
interest (ROI) shown in Fig.~\ref{fig:GC_flux}:  (i) $40^\circ \times 40^\circ$ with 
$1^\circ < |b| < 20^\circ$, $|l|<20^\circ$;  and (ii) full sky with $|b|>1^\circ$, 
where $b$ and $l$ are the Galactic latitude and longitude, respectively.
The authors of Ref.~\cite{Daylan:2014rsa} also perform an analysis for a third, 
smaller region $|b|<5^\circ$, $|l|<5^\circ$, but there are fewer statistics and thus 
a larger energy binning for this ROI.  As we later discuss, our model-independent 
analysis benefits from using regions with higher statistics.

\begin{figure}[t]
  \centering
  \includegraphics[width=8.7cm]{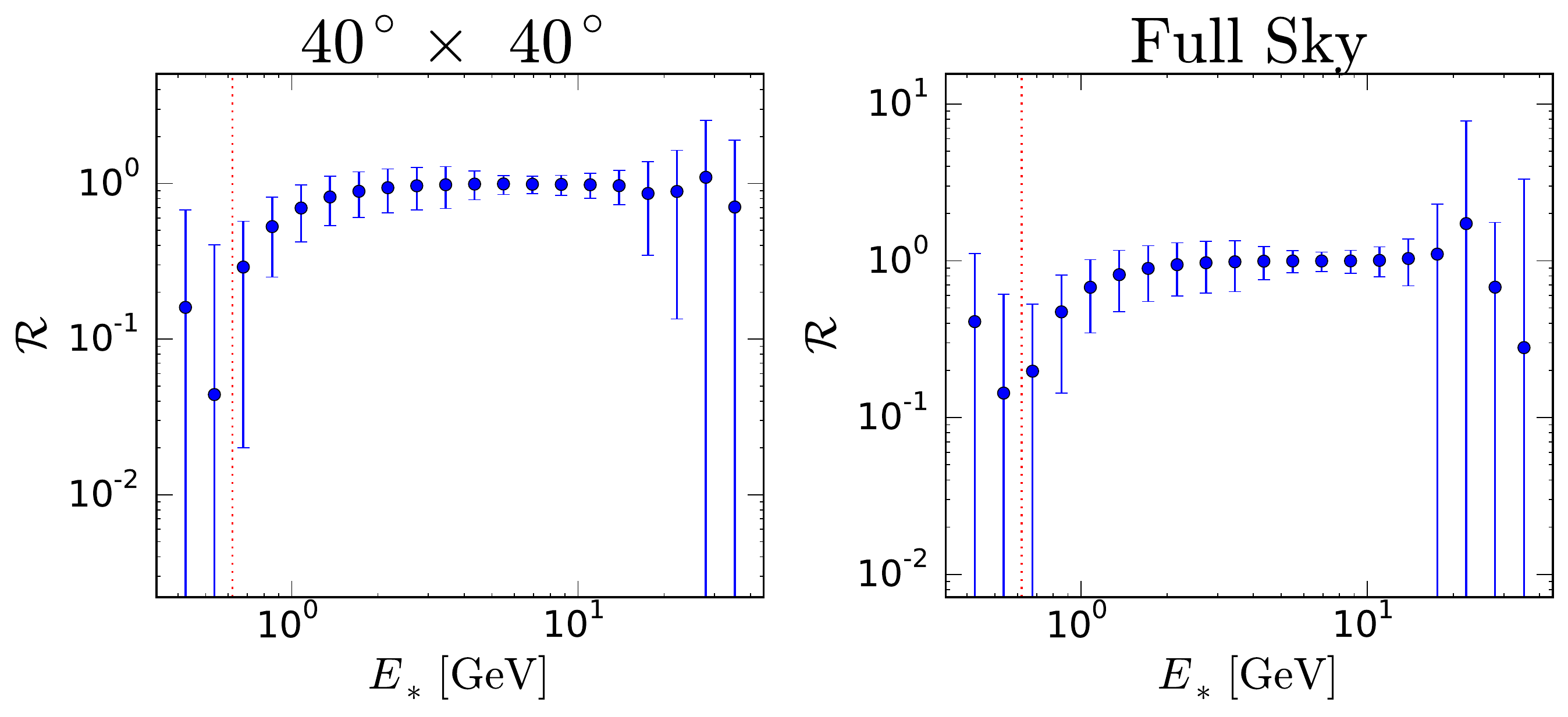}
  \caption{Testing the energy-duality hypothesis for the Galactic-Center gamma-ray excess.    
    We plot the asymmetric-to-symmetric ratio ${\cal R}(n_\ast)$ in 
    Eq.~(\protect\ref{eq:AsymOverSym}) as a function of the self-dual bin $n_\ast$ containing 
    $E_\ast$.  We find that ${\cal R}(n_\ast)$ is minimized at values near 
    ${\cal R}(n_\ast)\sim 10^{-1}-10^{-2}$ for $n_\ast=3$, signifying a
    reasonably good fit to a self-dual energy spectrum invariant under 
    $E_\gamma \to E_\ast^2/E_\gamma$ with a self-dual energy near $E_\ast \sim 0.5$~GeV.
    The superimposed red dashed lines indicate the best-fit values for $E_\ast$ taken 
    from Fig.~\ref{fig:fit} for the DDM model in Sect.~\ref{sec:Models}.
           }
  \label{fig:AsymOverSym}
\end{figure}

Our interest in this paper is in the overall shape of this gamma-ray flux excess ${\cal F}$, and 
in particular the possibility that this spectrum exhibits an energy-duality invariance
under $E_\gamma\to E_\ast^2/E_\gamma$ for some $E_\ast$.  Note that this is equivalent 
to $x\to 1/x$ where $x\equiv E_\gamma/E_\ast$, or $\log(x) \to -\log(x)$.  Thus,  
if this spectrum had an exact energy duality with $E_\ast$ logarithmically centered 
inside a particular energy bin $n_\ast$, the plot in Fig.~\ref{fig:GC_flux} would be 
completely symmetric on a logarithmic scale.  In order to test this hypothesis,
we quantify the extent to which this spectrum exhibits an energy duality by calculating 
the ratio of the asymmetric part versus the symmetric part of the spectrum as a function 
of the chosen reference bin $n_\ast$ with respect to which these symmetries are 
calculated:
\begin{equation}
 {\cal R}(n_\ast)
   ~\equiv~
    \displaystyle{  \frac{\sum_{n=1}^{n_\textrm{max}} |{\cal F}_{n_\ast+n} - {\cal F}_{n_\ast-n}|}
    {\sum_{n=1}^{n_\textrm{max}} |{\cal F}_{n_\ast+n} + {\cal F}_{n_\ast-n}|}}~.
    \label{eq:AsymOverSym}
\end{equation}
Here ${\cal F}_m$ is that portion of the excess differential flux residing within the $m^{\rm th}$ energy bin, 
and $n_\textrm{max} \equiv {\rm min}(n_{\ast}-1, 22-n_{\ast})$.  Our results are plotted 
in Fig.~\ref{fig:AsymOverSym} for $n_\ast= 2,...,21$.  The value of $n_\ast$ for which 
the asymmetric-to-symmetric flux ratio ${\cal R}(n_\ast)$ is minimized indicates that the 
spectrum is most consistent with an energy duality for which the self-dual energy $E_\ast$ 
is contained in that particular bin.  Of course, we do not expect a perfect energy duality 
to be evident in the data.  In particular, aside from statistical fluctuations, we do not 
expect a binned energy-dual spectrum to have a perfectly vanishing minimum 
asymmetric-to-symmetric ratio ${\cal R}(n_\ast)$, since an arbitrary binning will not
logarithmically center a particular bin on $E_\ast$.

%==============================================
%==============================================

The results in Fig.~\ref{fig:AsymOverSym} suggest that the GC excess may indeed be 
energy-dual with respect to the $n_\ast=3$ energy bin, corresponding to 
$E_\ast \sim 0.5~\gev$.  However, there are a few difficulties in confidently determining 
the value of $E_\ast$ even if this energy duality does in fact exist.  First, the energy 
binning prevents $E_\ast$ from being determined to within $\sim 125~\mev$, although any 
particular energy-dual model may yield a more precise fit.  Second, since the statistical 
variance in the differential flux at lower energies is large, the apparent energy duality 
may be an artifact of small statistics.  Both better energy resolution
and higher statistics are important in making a precise model-independent determination 
of $E_\ast$.  (We note in this context that observations from, \textit{e.g.}, 
GAMMA-400~\cite{Galper:2014pua} are expected to have a better resolution than the 
Fermi-LAT near 1~GeV.)  Third, we see that there are only two bins below that which 
contains $E_\ast$.  Thus data at even lower energies is needed in order to further test 
the energy duality from the higher-energy tail of the excess.  Satellites that can probe 
the $10~\mathrm{MeV} \lesssim E_\gamma \lesssim 1~\mathrm{GeV}$ energy range with 
sufficient resolution will be crucial in determining if the photon spectrum associated 
with the GC excess is indeed energy-dual.  Finally, we again stress that error bars 
in Figs.~\ref{fig:GC_flux} and \ref{fig:AsymOverSym} represent only statistical 
errors, and our discussion assumes that the  systematic uncertainties (which we have 
been ignoring) do not ruin the energy duality. 
Indeed, all of these issues must be borne in mind when attempting to draw any robust 
conclusions from the shape of the GC gamma-ray spectrum extracted from current Fermi 
data.  We shall return to these issues, and explore what such future experiments could 
do to foster a more robust claim for such a duality, in Sect.~\ref{sec:conclusion}.

Moreover, it should also be kept in mind that while annihilating/decaying dark matter 
provides one possible explanation for the excess observed in Fermi data, other explanations 
have been advanced as well.  For example, a population of unresolved milisecond pulsars has 
been advanced as a plausible explanation for this excess~\cite{Bartels:2015aea,Lee:2015fea}.
Alternatively, it has been pointed out that the GC excess can be explained in terms 
of leptonic cosmic-ray bursts of an astrophysical 
origin~\cite{Petrovic:2014uda,Cholis:2015dea}, once the effects of standard steady-state 
diffusion~\cite{Gaggero:2015nsa} are properly taken into account.

Assuming the GC excess does have a dark-matter origin and is indeed energy-dual, 
it is also nevertheless possible that certain spectral features are masked due to the 
finite energy binning.  As the photon energy approaches $E_\ast \sim 0.5~\gev$, the 
spectrum has a single, sharp peak and falls off (nearly) monotonically above and below 
$E_\ast$.  However, the spectrum could potentially consist of multiple overlapping 
peaks that cannot be resolved.  Indeed, such a scenario could still be energy-dual.  
Alternatively, the spectrum could exhibit a plateau or smooth bump instead of a cuspy 
peak, as long as the critical size needed to distinguish between these possibilities is 
smaller than the size of the energy bins.

In the following sections, we shall assume that the GC gamma-ray excess indeed exhibits 
an energy-dual spectrum and discuss the physical implications that such an observation 
might have in terms of annihilating and/or decaying dark matter.

%%%%%%%%%%%%%%%%%%%%%%%%%%%%%%%%%%%%%%%%%%%%%%%%%%%%%%%%%%%%%%%%%%%%%%%%%%%%%%%

\section{Boosts and Boxes:~  Building an Energy-Dual Spectrum\label{sec:Framework}}

%%%%%%%%%%%%%%%%%%%%%%%%%%%%%%%%%%%%%%%%%%%%%%%%%%%%%%%%%%%%%%%%%%%%%%%%%%%%%%%

In this section, we shall discuss the underlying kinematics that might lead to an 
energy-dual photon spectrum.  Our focus shall be on energy-dual spectra which resemble 
a single continuum ``bump'' --- {\it i.e.}\/, spectra whose magnitudes first rise as a function 
of energy and then fall.

%%%%%%%%%%%%%%%%%%%%%%%%%%%%%%%%%%%%%%%%%%%%%%%%%%%%%%%%%%%%%%%%%%%%%%%%%%%%%%%
\subsection{Filling boxes through boosts}
%%%%%%%%%%%%%%%%%%%%%%%%%%%%%%%%%%%%%%%%%%%%%%%%%%%%%%%%%%%%%%%%%%%%%%%%%%%%%%%

The most trivial example of a self-dual photon energy spectrum is a spectral line
corresponding to mono-energetic photons with $E_\gamma=E_\ast$.  Indeed, this spectrum 
is self-dual regardless of the spatial orientations of the various photon momenta,
and is thus self-dual if the photon momenta are distributed isotropically.  However, 
what is perhaps less trivial is that the energy spectrum of such photons remains self-dual 
even if such photons are boosted relative to the lab frame in which the photon energies 
are measured.  Indeed, all that is required is that the photons continue to be 
distributed isotropically in the boosted frame.  To see this, let us imagine that a 
given photon with energy $E_\ast$ is boosted with a velocity $\beta$, with an angle 
$\theta$ between the photon momentum and the boost direction.  In the lab frame, the 
corresponding photon energy will be given by
\beq
          E_\gamma ~=~ \gamma E_\ast (1+ \beta \cos\theta)~
\eeq
where $\gamma = (1-\beta^2)^{-1/2}$ is the usual relativistic factor.  Since the 
probability distribution for these photons is assumed to be isotropic in the boosted 
frame, all values of $\cos\theta$ are sampled with equal probability.  Thus, the 
resulting photon spectrum will fill out a spectral ``box'' in energy space stretching 
between $E^{\pm}_\gamma \equiv \gamma E_\ast (1\pm \beta)$. 
It is easy to verify that such a spectrum continues to be duality invariant.  For 
$\beta=0$ (vanishing boost), this box collapses to the original spectral line at 
$E_\gamma = E_\ast$.  However for non-zero boosts this spectral line expands in a 
self-dual way to form a box of width
\beq
        \Delta E ~\equiv~ E_\gamma^+ - E_\gamma^- ~=~ 2 \gamma \beta E_\ast~,
\eeq
logarithmically centered at 
$E_\ast=\sqrt{ E_\gamma^- E_\gamma^+}$.

Such a kinematics is easy to realize if our mono-energetic photons are isotropically 
emitted through the decay of a massive particle $\phi$ with momentum $p_\phi$.
In this case the momentum $p_\phi$ produces the required boost, whereupon we can 
identify $\gamma= E_\phi/m_\phi$ and $\gamma\beta= p_\phi/m_\phi$.  The width of the 
resulting spectral box is then given by
\beq
       \Delta E ~=~ 2 E_\ast {p_\phi\over m_\phi} ~=~ 
       {2 E_\ast\over m_\phi} \sqrt{ E_\phi^2 - m_\phi^2}~.
\label{eq:width}
\eeq
This width vanishes in the zero-boost limit $E_\phi\to m_\phi$.  Otherwise, the width 
of this box grows as a function of $E_\phi$ and encompasses an ever-increasing range of 
energies.  We can further ensure that the photons emitted through such a $\phi$ decay 
will be isotropic if $\phi$ is spinless or at least unpolarized; likewise such photons 
will be mono-energetic in the $\phi$ rest frame if this is a two-body decay, 
\ie,  $\phi\to \gamma Y$ for some particle $Y$.
In this case we find that
\beq
             E_\ast ~=~ {m_\phi^2 - m_Y^2 \over 2 m_\phi}~.
\eeq

Given this setup, we may ask what minimum boost (\ie, what minimum value of $E_\phi$) is 
required in order for our resulting photon spectrum in the lab frame to include
a given energy $E_\gamma$.  Clearly, this is tantamount to determining the minimum 
value of $E_\phi$ for which $E_\gamma^- \leq E_\gamma\leq E_\gamma^+$.  Solving these 
inequalities, we find that we must have
\beq
            E_\phi ~\geq~ {m_\phi\over 2} \left( x+ {1\over x}\right)~~~~~ 
            {\rm where}~~ x\equiv {E_\gamma \over E_\ast}~.
\label{minboost}
\eeq
This result displays the expected energy-duality invariance under $x\to 1/x$, and
thus holds regardless of whether $E_\gamma< E_\ast$ or $E_\gamma> E_\ast$.
Moreover, as expected, we see that no boost at all is required if $E_\gamma=E_\ast$:
indeed for $x=1$ we find from Eq.~(\ref{minboost}) that any $E_\phi \geq m_\phi$ will 
suffice.

%%%%%%%%%%%%%%%%%%%%%%%%%%%%%%%%%%%%%%%%%%%%%%%%%%%%%%%%%%%%%%%%%%%%%%%%%%%%%%%
\subsection{Stacking boxes to build an energy-dual spectrum}
%%%%%%%%%%%%%%%%%%%%%%%%%%%%%%%%%%%%%%%%%%%%%%%%%%%%%%%%%%%%%%%%%%%%%%%%%%%%%%%

\begin{figure}[t]
  \centering
  \includegraphics[width=6.0cm]{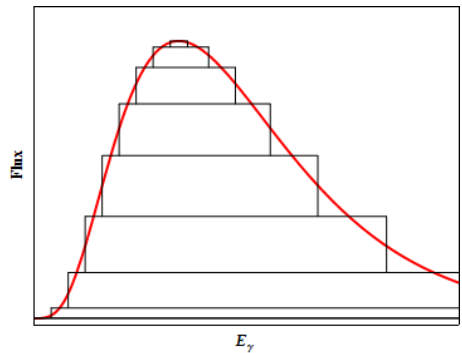}
  \caption{Stacking boxes (black outlines) to build a self-dual photon energy spectrum (red curve).
    By stacking self-dual boxes of suitably chosen widths and heights, we can reproduce
     any bump-like self-dual photon energy spectrum.
      As described in the text, a given collection of boxes with various widths and heights
         corresponds to a specific differential number $dN_\phi/dE_\phi$
         of parent $\phi$ particles with different boost energies $E_\phi$.
           It is these parent particles
         whose two-body decays produce
        the photons whose bump-like energy spectrum  we are modeling.  }
  \label{fig:stacking}
\end{figure}

Thus far, we have seen that any massive particle $\phi$ that decays isotropically
into a two-body final state including at least one photon will lead to a self-dual 
``box''-like photon energy spectrum.  However, given this, it is not hard to imagine 
how we might realize a given self-dual ``bump''-like energy spectrum:   we simply 
stack different boxes on top of each other, utilizing boxes with suitably chosen 
widths and heights.    Indeed, any self-dual bump-like spectrum can be decomposed
into a collection of such boxes, in much the same way as any periodic curve can be 
Fourier-decomposed into cosines and sines of different frequencies.  This stacking 
procedure is illustrated in Fig.~\ref{fig:stacking}.

%=======================================================================================
%=======================================================================================

At a physical level, this procedure may be interpreted kinematically as follows.
As we have seen, a given box represents the energy spectrum of a photon emerging from 
the two-body decay of a massive particle $\phi$ with a given boost energy $E_\phi$: 
the width of the box corresponds to the boost energy $E_\phi$ via Eq.~({\ref{eq:width}), while the height of the 
box is determined by the (differential) number of such $\phi$ particles with that boost 
energy.   A given collection of boxes with various widths and heights therefore 
corresponds to a specific (differential) number $N_\phi$ of $\phi$ particles as a 
function of boost energy $E_\phi$.

Mathematically, if our bump-like photon spectrum corresponds to a differential photon 
number $dN_\gamma/dE_\gamma$, the process of superposition in Fig.~\ref{fig:stacking} 
corresponds to writing
\beq
  \frac{dN_\gamma}{dE_\gamma} ~=~ n_\gamma \int_{m_\phi}^\infty dE_\phi
    \, \frac{dN_\phi}{dE_\phi}
    \, \frac{\Theta(E_\gamma^+ - E_\gamma) \Theta(E_\gamma - E_\gamma^-)}
       {2 E_\ast {p_\phi/  m_\phi}} ~
\label{integrate}
\eeq
where $dN_\phi/ dE_\phi$ represents the corresponding differential number of decaying 
$\phi$ particles.
Indeed, as described above, we may realize any self-dual function
$dN_\gamma/dE_\gamma$ in Eq.~(\ref{integrate}) through an appropriate choice of 
$dN_\phi/dE_\phi$, provided that $dN_\gamma/dE_\gamma$ is truly ``bump-like'', 
decreasing monotonically away from its maximum in either direction with no smaller peaks 
elsewhere.
In Eq.~(\ref{integrate}), the Heaviside theta-functions in the numerator of the integrand
enforce the upper and lower energy limits of each box, while the width $\Delta E$ in the
denominator provides a proper corresponding normalization.  Finally, the quantity 
$n_\gamma$ denotes the number of mono-energetic photons produced per $\phi$ decay. 
For the process $\phi \rightarrow \gamma Y$ we have $n_\gamma=1$ unless $Y=\gamma$, 
in which case $n_\gamma=2$.

Note that Eq.~(\ref{integrate}) may equivalently be written as
\beq
  \frac{dN_\gamma}{dx} ~=~  n_\gamma  \int_{\frac{m_\phi}{2} (x+1/x)}^\infty dE_\phi
    \left[ \frac{dN_\phi}{dE_\phi} \, \frac{m_\phi}{2\sqrt{E_\phi^2 - m_\phi^2}}
    \right]~
  \label{eq:GenSpectrum}
\eeq
where $x \equiv E_\gamma / E_\ast$ and where the lower limit of integration comes from Eq.~(\ref{minboost}).
Of course, in writing these integrals we are assuming an essentially ``continuous'' 
collection of boxes, as would be required in order to produce a net photon energy 
spectrum which rises and falls smoothly compared with a corresponding detector 
resolution/binning.  Note that the integral in Eq.~\eqref{eq:GenSpectrum} depends on $x$ 
only through the lower limit of integration and the integrand is non-negative; thus, 
$dN_\gamma / dx$ decreases as the lower limit of integration increases. However, $x + 1/x$ 
is minimized at $x=1$ and grows monotonically as $x$ departs from $1$ in either direction. 
Thus, the spectrum $dN_\gamma / dx$ is maximized at $x=1$, and decreases monotonically as 
$x$ either increases or decreases away from this limit.

In stacking our boxes, it is interesting to distinguish between three distinct cases:
those which lead to a cusp for $dN_\gamma/dE_\gamma$
at the self-dual energy $E_\gamma=E_\ast$,
those which lead to a smoothly rounded maximum, and those 
which lead to a flat plateau.
These cases can be distinguished by examining the derivative 
of the differential photon number as we approach the self-dual energy $E_\ast$:
\begin{equation}
  \left. \frac{d^2 N_\gamma}{dx^2} \right|_{x \rightarrow 1} ~=~  {\rm sgn}(1-x)
            ~n_\gamma~ \frac{m_\phi}{2}
   \left. \frac{dN_\phi}{dE_\phi}  \right|_{E_\phi \to m_\phi} .~~
\end{equation}
Thus, if $dN_\phi / dE_\phi$ is non-vanishing as $E_\phi \to  m_\phi$, then the 
derivative of the spectrum is discontinuous, implying that $dN_\gamma / dx$ has a 
cuspy peak at $x=1$~\cite{KaustubhEnergyPeak,Chen:2014oha}.  By contrast, if 
$dN_\phi / dE_\phi$ approaches zero smoothly as $E_\phi \rightarrow m_\phi$, then the 
peak is a smooth bump.  However, if $dN_\phi / dE_\phi$ vanishes below a threshold 
energy $\overline{E} > m_\phi$, then the photon spectrum exhibits a plateau along its maximum.  In this 
case, $dN_\gamma / dx$ is constant for 
$x + 1/x < 2 \overline{E} / m_\phi$~\cite{KaustubhEnergyPeak,Stecker,Kim:2015gka}.
Further discussions concerning these observations can be found in 
Refs.~\cite{KaustubhEnergyPeak, Agashe:2012fs, Agashe:2013eba, Chen:2014oha,Agashe:2015wwa, 
Agashe:2015ike} within the context of collider phenomenology,
and in Refs.~\cite{DoojinAstroEnergyPeak, Kim:2015gka} within the context of 
gamma-ray astrophysics.  Of course, while the current data describing the GC excess is 
consistent with a sharp peak, it is also consistent with a narrow plateau or smoothly rounded maximum, provided the width of the bump is less than the resolution of the binning.

We thus conclude that an energy-dual gamma-ray spectrum can easily emerge if these 
photons result from the isotropic two-body decays of massive particles $\phi$ with a boost 
(or injection) spectrum $dN_\phi/dE_\phi$.  In such cases the peak (or center of a 
plateau or smooth bump region) of our photon distribution determines $E_\ast$, while the 
spectral shape encodes the boost (or injection) spectrum $dN_\phi/dE_\phi$.  Our remaining 
task, then, is to find a dark-matter model in which such an injection spectrum emerges 
naturally and is ultimately consistent with the GC excess.

%%%%%%%%%%%%%%%%%%%%%%%%%%%%%%%%%%%%%%%%%%%%%%%%%%%%%%%%%%%%%%%%%%%%%%%%%%%%%%%

\section{Dynamical Dark Matter and the Galactic-Center Excess\label{sec:Models}}

%%%%%%%%%%%%%%%%%%%%%%%%%%%%%%%%%%%%%%%%%%%%%%%%%%%%%%%%%%%%%%%%%%%%%%%%%%%%%%%

In principle, one can imagine many models of dark-sector physics in which dark-matter 
annihilations or decays produce a particle $\phi$ whose subsequent decays produce the 
photons which are observed emanating from the Galactic Center.  Likewise, there are 
many possibilities which give rise to a non-trivial injection spectrum $dN_\phi/dE_\phi$ 
for these intermediary particles.  For example, dark-matter decays or annihilations 
involving $N$-body final states with $N>2$ will lead to a non-trivial injection spectrum 
$dN_\phi/dE_\phi$ if $\phi$ is one of the resulting decay products.  Other more 
complicated scenarios are also possible.

One particularly simple possibility, however, is to imagine that each of the ``boxes'' 
discussed in Sect.~\ref{sec:Framework} corresponds to a different dark-matter particle 
$\chi_n$ in the dark sector.  Each $\chi_n$ can then decay or annihilate, producing a 
pair of intermediary particles $\phi$ which subsequently decay into two photons.
This kinematics is sketched in Fig.~\ref{fig:model}.  In the limit in which each 
$\chi_n$ is non-relativistic with respect to the lab (observer) frame, the 
intermediaries $\phi$ resulting from each such annihilation or decay will be generated 
with a fixed boost whose magnitude depends on the mass of $\chi_n$.  Thus, if we wish 
to construct dark-matter models based on the kinematic configurations shown in 
Fig.~\ref{fig:model}, we are naturally led to consider dark sectors comprising 
different dark-matter particles $\chi_n$ of different masses.

Remarkably, this is precisely one of the ingredients of the Dynamical Dark Matter 
(DDM) framework~\cite{DDM1,DDM2}.  In general, DDM models contain multiple dark-matter 
components which together form an ensemble whose phenomenological viability is the 
result of a balancing between decay widths and relic abundances across the ensemble.
If the mass splitting between the ensemble components is smaller than the energy resolution 
of the detector in question, the boost distribution of the intermediary particles $\phi$ 
appears continuous and thus the resulting photon spectrum appears as a continuum bump.
A similar idea has been adopted in Ref.~\cite{MeVDDM} within the context of MeV-range 
gamma-ray detection experiments.

%%%%%%%%%%%%%%%%%%%%%%%%%%%%%%%%%%%%%%%%%%%%%%%%%%%%%%%%%%%%%%%%%%%%%%%%%%%%%%%
\subsection{Constructing a DDM model \label{modelsect}}
%%%%%%%%%%%%%%%%%%%%%%%%%%%%%%%%%%%%%%%%%%%%%%%%%%%%%%%%%%%%%%%%%%%%%%%%%%%%%%%

Towards this end, we therefore consider a DDM model in which a (potentially) large 
number of DM components $\chi_n$  form a dark-matter ensemble.  We label the DDM 
components by the index $n=0,...,N$
in order of increasing mass. 
We assume that each $\chi_n$ has a relic abundance $\Omega_n$ such that the ensemble as 
a whole carries the observed total dark-matter relic abundance.  Indeed, such DDM ensembles are 
realized in various well-motivated physics models beyond the SM, including scenarios 
with extra spacetime dimensions~\cite{DDM1,DDM2,Dienes:2012jb}, confining hidden-sector 
gauge groups~\cite{DDMmodel1}, large spontaneously-broken symmetry 
groups~\cite{DDMmodel2,DDMmodel3}, and even certain string 
configurations~\cite{DDMmodel1,anupam}.

%======================================
\begin{figure}[t]
  \centering
  \includegraphics[width=8.7cm]{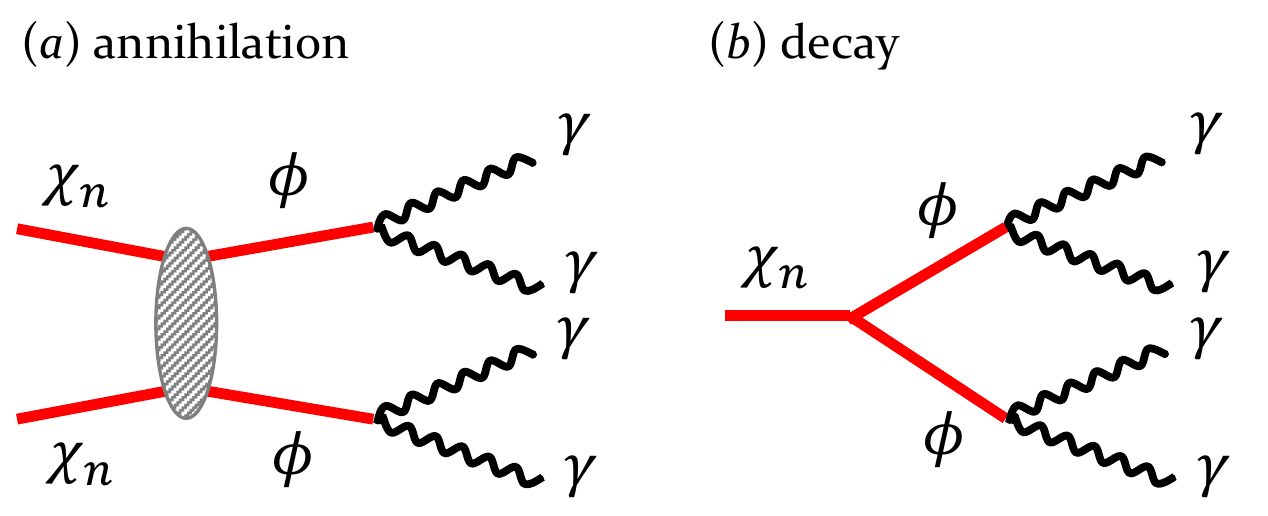}
  \caption{Annihilating and decaying DDM model scenarios under consideration.
    The DDM components $\chi_n$ annihilate or decay into the same intermediary 
    particles $\phi$, which subsequently decay to two photons.}
  \label{fig:model}
\end{figure}
%======================================

In the case of DM annihilation, we consider a pair of $\chi_n$'s that annihilate
to two $\phi$'s, each of which subsequently decays into two photons, as shown in 
Fig.~\ref{fig:model}(a).  (For simplicity, we shall not consider the possibility of 
coannihilation by $\chi_m$ and $\chi_n$ where $m\neq n$.)  As an example, $\chi_n$ 
could be a Dirac fermion and $\phi$ a singlet pseudoscalar (\textit{e.g.}, a 
``dark pion'' or an axion-like particle).  A possible Lagrangian would then take the 
form
\begin{equation}
  \mathcal{L}_{\textrm{ann}} ~\ni~ \sum_{n=0}^N \frac{c_n}{\Lambda}
  \bar{\chi}_n\chi_n \phi\phi+\frac{1}{f_{\phi}}\phi
  F_{\mu\nu}\tilde{F}^{\mu\nu}~,
\end{equation}
where $\tilde{F}^{\mu\nu}$ denotes the usual dual field strength tensor, $\Lambda$ 
is the scale of the effective field theory that governs DM annihilations, and $f_{\phi}$ 
is the symmetry-breaking scale that gives mass to the axion-like particle.  The coupling 
$c_n$ between $\phi$ and $\chi_n$ generically differs from component to component and so 
is also indexed by $n$.

In the case of DM decay, by contrast, we imagine a similar process in which a single 
$\chi_n$ decays into two $\phi$'s, {\it i.e.}\/, 
$\chi_n\rightarrow 2\phi \rightarrow 4\gamma$, as shown in Fig.~\ref{fig:model}(b). 
One possible scenario involves a scalar $\chi_n$ that decays into a pair of singlet 
pseudoscalar $\phi$ particles.  The possible corresponding Lagrangian would then take 
the form
\begin{equation}
  \mathcal{L}_{\textrm{dec}} ~\ni~ \sum_{n=0}^N c_n'M \chi_n\phi\phi
  +\frac{1}{f_{\phi}}\phi F_{\mu\nu}\tilde{F}^{\mu\nu}~,
  \label{eq:CouplingLagrangian}
\end{equation}
where $M$ is an associated mass scale which depends on the details of the underlying model.
As in the annihilation case, the coupling $c'_n$ between $\phi$ and $\chi_n$ can
be different for different components.

However, this is not all.
DDM models do not merely have a random assortment of dark-matter components --- these 
components must also have properties such as masses, abundances, and decay widths which obey specific 
{\it scaling relations}\/.  These scaling relations emerge naturally from a variety of 
underlying DDM constructions~\cite{DDM1,DDM2,Dienes:2012jb, DDMmodel1, DDMmodel2,DDMmodel3}.
The question that remains, then, is not merely whether there exists a non-trivial 
intermediary injection spectrum $dN_\phi/dE_\phi$ that can fit the GC excess,
{\it but whether this injection spectrum is also consistent with 
an underlying dark sector whose individual components exhibit scaling relations of the 
sort DDM assumes.}\/  Only this would be the true test of an underlying DDM-based origin 
for the GC excess.

We therefore consider how physical quantities such as the relic abundances, cross sections, 
and decay widths associated with our dark-matter components $\chi_n$ vary across the ensemble.
In general, these quantities can be parametrized in terms of the corresponding masses $m_n$.
As a result, the photon flux $\Phi_n$ associated with $\chi_n$ (resulting from either decay
or annihilation) will also depend on the mass 
$m_n$.  For concreteness, just as in Ref.~\cite{MeVDDM}, we shall consider the case where 
$\Phi_n$ scales with $m_n$ according to a simple power law of the form
\begin{equation}
  \Phi_n ~=~ \Phi_0\left(\frac{m_n}{m_0} \right)^{\xi}
  ~=~ \Phi_0\left(\frac{\sqrt{s_n}}{\sqrt{s_0}} \right)^{\xi}~,
\label{eq:scalingbeh}
\end{equation}
where the scaling exponent $\xi$ is taken to be a free parameter.  Note that we replace 
the masses in Eq.~(\ref{eq:scalingbeh}) by the center-of-mass (CM) energies $\sqrt{s_n}$ 
in order that our results are expressed in a form applicable to both the annihilation 
and decay scenarios.  In the non-relativistic regime, $\sqrt{s_n}$ is equal to $2m_n$ 
or $m_n$ for annihilating or decaying DM models, respectively.

In most DDM models, the masses $m_n$ can typically be parametrized in terms of
the mass $m_0$ of the lightest DDM component,
a mass-splitting parameter $\Delta m$, and
a scaling exponent $\delta$:
\begin{equation}
  m_n ~=~ m_0 + n^{\delta}\Delta m ~.
\end{equation}
The CM energy gap between neighboring DM states, {\it i.e.}\/, $\Delta(\sqrt{s_n})\equiv \sqrt{s_{n+1}}-\sqrt{s_n}$, is simply given by
\begin{equation}
  \Delta(\sqrt{s_n}) ~=~ \left\{
  \begin{array}{r l}
    2\left[(n+1)^{\delta}-n^{\delta} \right]\Delta m & \text{ for annihilation} \\ [2mm]
    \left[(n+1)^{\delta}-n^{\delta} \right]\Delta m & \text{ for decay~,}
    \end{array}\right.~
  \label{eq:delsn}
\end{equation}
which is valid up to $n=N-1$.
In this paper, we shall choose $\delta=1$, as arises in cases where the DDM ensemble 
are the states in a Kaluza-Klein tower.  With this choice of $\delta$, the mass spectrum 
of the DDM ensemble has a uniform spacing.  This allows us to write the CM energy gap as 
$\Delta(\sqrt{s})$ and thereby eliminate the unnecessary subscript $n$.  Indeed, we find 
that $\Delta(\sqrt{s})$ is $2\Delta m$ for annihilation and $\Delta m$ for decay.

Given this scaling behavior, we can now 
calculate the differential photon number $dN_\gamma/dE_\gamma$ corresponding to our DDM ensemble.
To do this, we shall 
work in the continuum limit in 
which $\Delta m\rightarrow 0$.  In this limit, we no longer have a discrete set of
energies $\sqrt{s_n}$;
we instead have a continuous CM energy $\sqrt{s}$ stretching
between $\sqrt{s_0}$ and $\sqrt{s_N}$.
Indeed, we may replace sums $\sum_{n=0}^N$ with integrals $\int_{\sqrt{s_0}}^{\sqrt{s_N}} d\sqrt{s}/\Delta(\sqrt{s})$.
Likewise, we no longer have a discrete set of
individual contributions $\Phi_n$ to the total flux at different discrete values of
$\sqrt{s_n}$;  we instead have a {\it function}\/ $\Phi(\sqrt{s})$ which describes the
total flux emerging from an underlying dark-matter annihilaton or decay
with CM energy $\sqrt{s}$.
In other
words, in this limit, Eq.~(\ref{eq:scalingbeh}) becomes
\begin{equation}
   \Phi(\sqrt{s}) ~=~
  \Phi_0\left(\frac{\sqrt{s}}{\sqrt{s_0}} \right)^{\xi} ~.
\label{eq:scalingbehcont}
\end{equation}
Note that since $\Phi(\sqrt{s})$ is not a {\it differential}\/ flux,
it carries no spectral information about the resulting
photons.  Rather, this quantity represents a particular contribution to
the {\it total}\/ gamma-ray flux.

While there are many ways in which we might calculate the total flux
corresponding to our DDM ensemble, we shall here follow a somewhat
quick and intuitive path which is similar in spirit to the ``stacking boxes''
discussion above.  A more rigorous derivation leading to the same result
(and justifying its overall normalization) appears in the Appendix.

We shall assume that each ensemble constituent 
$\chi_n$ annihilates or decays into a pair of $\phi$ particles, each with energy 
$E_\phi=\sqrt{s_n}/2$, and that each such $\phi$ particle in turn decays into a pair of photons.  
Thus, the differential number of $\phi$ particles produced by dark-matter annihilation or decay with
energy $E_\phi$ is proportional to the total flux density at $\sqrt{s} = 2E_\phi$:
\bea
  \frac{dN_\phi}{dE_\phi} ~\propto~ \Phi(2E_\phi) ~.
\label{weightings}
\eea
Next, we recognize that for each ensemble component $\chi_n$,  
the corresponding contribution to $dN_\phi/dE_\phi$ may be written as 
\beq
        {dN_\phi^{(n)}\over dE_\phi}~=~ 2 \, \delta\left( E_\phi - \sqrt{s_n}/2\right)~
\label{lightest}
\eeq
where the prefactor indicates that there are precisely two $\phi$ particles 
produced from the annihilation/decay of each $\chi_n$.
Thus, just as we stack boxes with appropriate heights in order to build our total spectrum
as in Fig.~\ref{fig:stacking},
we can sum over all of the states in the ensemble with the appropriate weightings given in 
Eq.~(\ref{weightings}) in order to  build our total ``effective'' 
differential number 
$dN_\phi/dE_\phi$: 
\beqn
    {dN_\phi\over  dE_\phi} &\propto&
        2\, \int_{\sqrt{s_0}}^{\sqrt{s_N}} {d\sqrt{s}\over \Delta (\sqrt{s})} \, 
              \left( {\sqrt{s}\over \sqrt{s_0}}\right)^\xi
            \delta(E_\phi- \sqrt{s}/2)~~~~~~~\nonumber\\
         &=& {4\over \Delta(\sqrt{s})} \left({2E_\phi\over\sqrt{s_0}}\right)^\xi ~\nonumber\\ 
            && ~\times \Theta \left(\frac{\sqrt{s_N}}{2} -E_\phi \right)
          \Theta \left(E_\phi - \frac{\sqrt{s_0}}{2} \right)~.~~~~
  \label{eq:dNdEphi}
\eeqn
Indeed, this is precisely the injection spectrum $dN_\phi/dE_\phi$
which has appeared throughout the main body of this paper thus far.
This notion of an ``effective'' $dN_\phi/ dE_\phi$ will be discussed
more precisely in the Appendix.

Note that the behavior of the injection spectrum $dN_\phi/dE_\phi$ as 
$E_\phi \rightarrow m_\phi$ depends on $\xi$ and $s_0$.  For $s_0 > 4m_\phi^2$, this 
injection spectrum is identically zero in the range $m_\phi < E_\phi < \sqrt{s_0}/2$, 
yielding a plateau-like maximum in the photon spectrum.  On the other hand, for 
$s_0 \sim 4m_\phi^2$, the injection spectrum decreases as $E_\phi \rightarrow m_\phi$ 
for positive $\xi$, yielding only a mild discontinuity in the derivative of the 
resulting photon spectrum near its maximum.  
For negative $\xi$, however, one would find a sharper peak.

%============================================
\begin{figure*}[t]
  \centering
  \includegraphics[width=8.4cm]{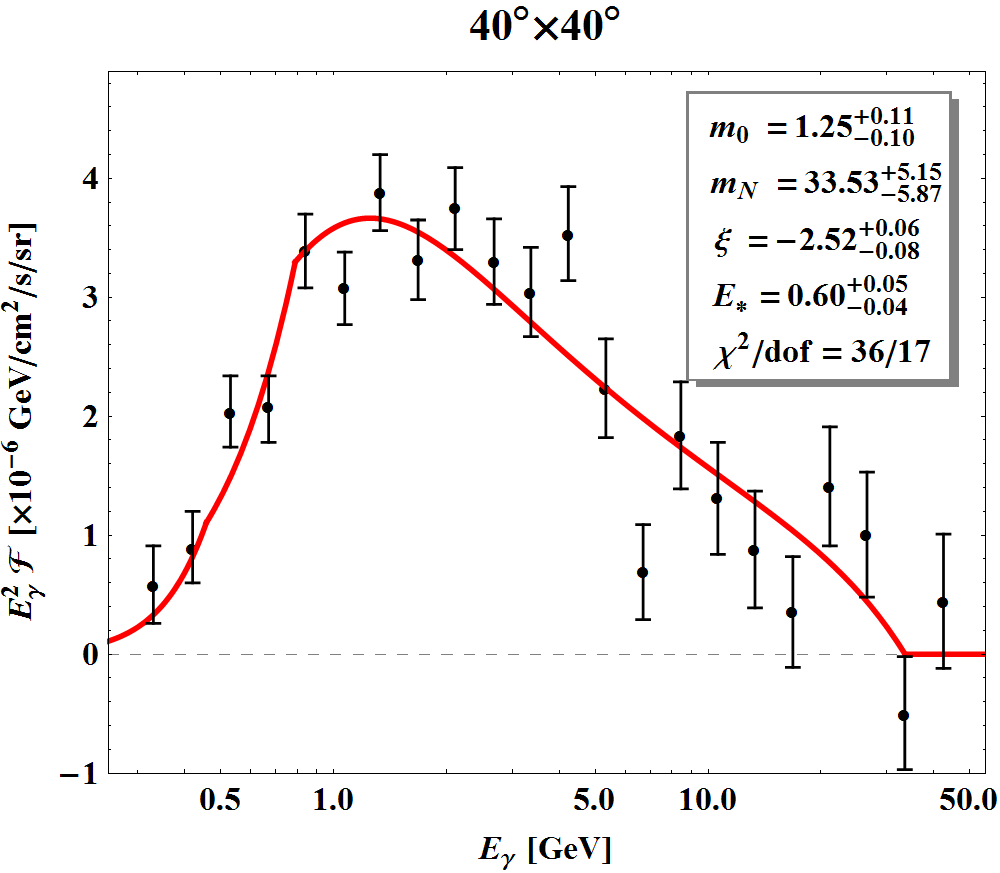}\hspace{0.5cm}
  \includegraphics[width=8.4cm]{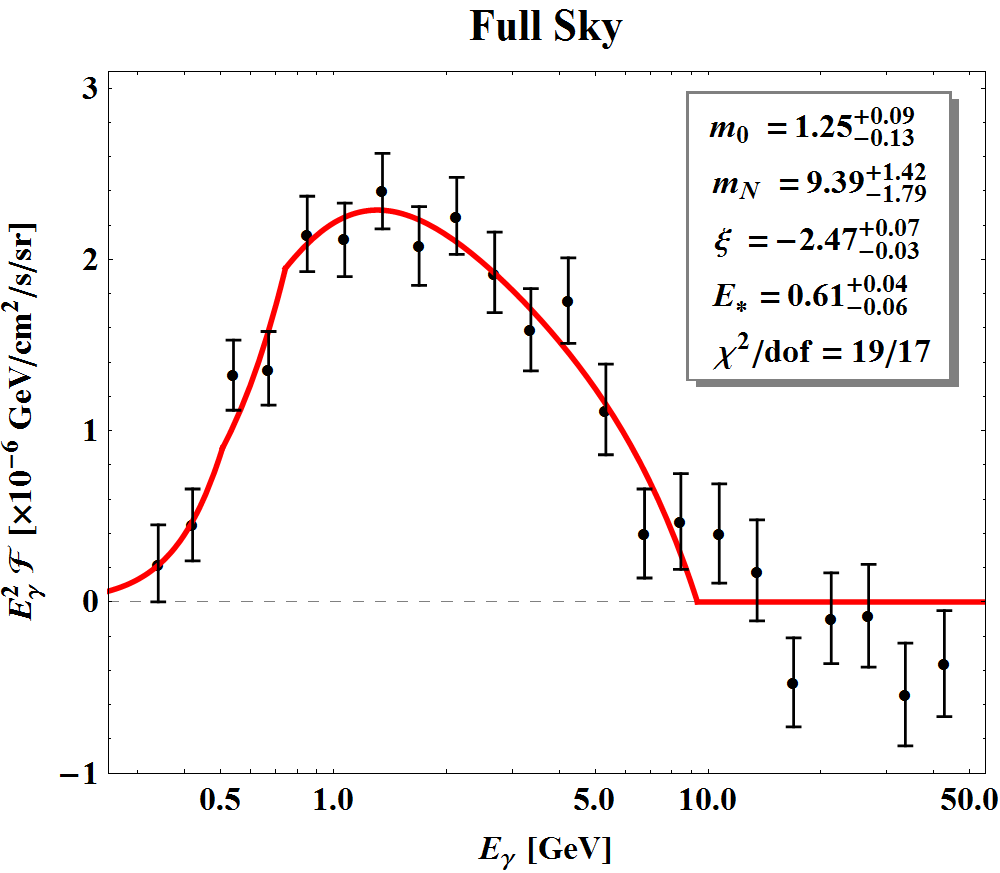}
  \caption{The GC photon-excess spectra (black dots and error bars) extracted from 
    Ref.~\protect\cite{Daylan:2014rsa}, corresponding to ROI's~(i) and (ii) for the 
    left and right panels respectively, with the best-fit DDM flux superimposed in red.
    Input parameters for these best-fit curves are also shown in each panel, with the best-fit values
       for $m_0$, $m_N$, and $E_\ast$ quoted in GeV.  These 
    results indicate that our DDM model is successful in modelling the GC 
    photon flux excess.  }
\label{fig:fit}
\end{figure*}
%============================================

Evaluating the photon spectrum that follows from 
this result for $dN_\phi/dE_\phi$ 
is now simply a matter of substituting Eq.~(\ref{eq:dNdEphi}) into Eq.~(\ref{eq:GenSpectrum})
with $n_\gamma =2$.  We thus obtain
\begin{eqnarray}
  \frac{dN_\gamma}{dx} &\propto & 
     {2m_\phi \over \Delta(\sqrt{s})} \left({2m_\phi\over \sqrt{s_0}}\right)^\xi \nonumber\\
    & & \times \left[B_{z_+}\left(-\frac{\xi}{2},\frac{1}{2}\right) 
                   - B_{z_-}\left(-\frac{\xi}{2},\frac{1}{2}\right)\right]~,~~~~~~~
  \label{eq:expectedE}
\end{eqnarray}
where $B_z(a,b)$ denotes the incomplete Euler beta function and where 
\begin{eqnarray}
  z_+ &\equiv& \max \left(\frac{4m_\phi^2}{s_N}, \min\left[\frac{4m_\phi^2}{s_0},
  \,\frac{4}{(x+1/x)^2}\right] \right)~,\nonumber\\
  z_- &\equiv& \frac{4m_\phi^2}{s_N} ~.~~~~~
  \label{eq:norm}
\end{eqnarray}

Our final step is to convert this expression for $dN_\gamma/dx$ 
into a total differential {\it flux}\/ $d\Phi/dE_\gamma$.
However, in terms of the total ``effective'' $dN_\gamma/dE_\gamma$ given in Eq.~(\ref{eq:expectedE}),
we know that
\beq
     {{ d\Phi/dE_\gamma} \over \Phi_0} ~=~ 
            {dN_\gamma/dE_\gamma \over N_\gamma^{(0)}} 
            ~=~ {dN_\gamma/dE_\gamma \over 4}~. 
\label{ratios}
\eeq
It then follows that
\beq
     { d\Phi\over dE_\gamma} ~=~  {\Phi_0\over 4} \, {dN_\gamma\over dE_\gamma} ~=~
      {\Phi_0\over 4E_\ast} \, {dN_\gamma\over dx} ~,
\label{fluxx}
\eeq
where $dN_\gamma/dx$ is given in Eq.~(\ref{eq:expectedE}).

The result in Eq.~(\ref{fluxx}) is sufficient for understanding the shape of the overall photon spectrum.
The normalization of this spectrum nevertheless remains unfixed because of the unknown normalization in
Eq.~(\ref{eq:expectedE}).
In general, the derivation we have provided is not capable of determining the correct normalization.
However, given the prefactors already present in Eq.~(\ref{eq:expectedE}), we shall see 
in the Appendix that the 
remaining constant of proportionality in Eq.~(\ref{eq:expectedE}) is actually equal to one.
Thus, in what follows, we shall feel free to replace the proportionality sign in 
Eq.~(\ref{eq:expectedE}) with an equals sign.

%%%%%%%%%%%%%%%%%%%%%%%%%%%%%%%%%%%%%%%%%%%%%%%%%%%%%%%%%%%%%%%%%%%%%%%%%%%%%%%
\subsection{Fitting the observed excess}
%%%%%%%%%%%%%%%%%%%%%%%%%%%%%%%%%%%%%%%%%%%%%%%%%%%%%%%%%%%%%%%%%%%%%%%%%%%%%%%

Given the expression 
in Eq.~(\ref{fluxx})
for the differential flux predicted by our DDM model,
we can now perform a fit to the spectrum of the GC excess observed in the 
Fermi-LAT data.  
Note that the data reported in Ref.~\cite{Daylan:2014rsa} 
is actually quoted in terms of 
the rescaled differential flux 
$E_\gamma^2 d^2\Phi/(dE_\gamma d\Omega) \equiv E_\gamma^2 {\cal F}$.
Thus, putting the pieces together,
we shall therefore fit this data 
to the predicted DDM template function
\beqn
   E_\gamma^2 {\cal F} &=&  
     ~{ E_\gamma^2 \over \Delta \Omega} ~ \Xi\, \left( {4E_\ast\over \sqrt{s_0}}\right)^\xi \nonumber\\
     && ~~\times \left[B_{z_+}\left(-\frac{\xi}{2},\frac{1}{2}\right) - 
                B_{z_-}\left(-\frac{\xi}{2},\frac{1}{2}\right)\right]~,~~~~~~~~~
\label{templatefct}
\eeqn
where 
\begin{eqnarray}
  z_+ &\equiv& \max \left(\frac{16E_\ast^2}{s_N}, \min\left[\frac{16 E_\ast^2}{s_0},
  \,\frac{4}{(E_\gamma/E_\ast + E_\ast/E_\gamma)^2}\right] \right)~\nonumber\\
  z_- &\equiv& \frac{16 E_\ast^2}{s_N} ~
  \label{eq:norm2}
\end{eqnarray}
and where
\beq 
           \Xi ~\equiv~ {\Phi_0\over \Delta(\sqrt{s})}~.
\eeq
Note that within this template we have replaced $m_\phi$ in favor of the self-dual energy $E_\ast$.
Moreover, because it provides a better fit to the spatial morphology of the excess,
we shall focus on the annihilating dark-matter case, for which $\sqrt{s_n} = 2m_n$.  Thus,
we take $\lbrace m_0, m_N, \xi, E_\ast, \Xi\rbrace$  as the five free parameters to which
we perform our fit. 
Since the first four parameters describe the underlying
particle-physics model, we would expect similar best-fit values to emerge for both ROI's.
By contrast, the normalization factor $\Xi$ depends not only on our specific particle-physics model
but also on astrophysical information
about the particular ROI --- information encapsulated by 
the corresponding ${\cal J}\/$-factor which is implicit
within $\Phi_0$.  For this reason, the best-fit values of $\Xi$ for our 
two ROI's need not be the same as  each other.

In our analysis of the data for both ROI's, we perform our fits using the 
standard $\chi^2$ statistic as our measure of goodness of fit.  Our best-fit results 
for ROI~(i) and ROI~(ii) are displayed in the left and right panels of Fig.~\ref{fig:fit},  
respectively.  The corresponding central values (black dots) for the gamma-ray flux
in each bin are also shown in each panel, along with their associated error bars.
The results shown in Fig.~\ref{fig:fit} indicate that our signal model reproduces the 
observational data for both ROI's rather well.  Indeed, the $\chi^2$ values for 
the fits performed on the data from ROI~(i) and ROI~(ii) are 36 and 19, respectively,
with 17 degrees of freedom in each case (as there are 22 data points 
for each of our five-parameter fits).  These numbers indicate that this DDM scenario is indeed
successful in accounting for the GC excess.

The best-fit values for the parameters $m_0$, $m_N$, $\xi$, and $E_\ast$ are shown 
in Fig.~\ref{fig:fit}, with the first, second, and fourth of these quantities quoted in 
GeV.~    All of the reported errors are given at the 68\% confidence 
level.  We also find that
\beq
    \Xi ~=~\left\{
    \begin{array}{l l}
       \phantom{3}1.81^{+0.15}_{-0.17} \times 10^{-6} ~({\rm GeV}\, {\rm cm}^{2} \, {\rm s})^{-1}
         & \hbox{ for ROI (i)} \\ [2mm]
       30.26^{+2.80}_{-2.85} \times 10^{-6} ~ ({\rm GeV}\, {\rm cm}^{2} \, {\rm s})^{-1}
         & \hbox{ for ROI (ii).}
   \end{array}\right.
\eeq
With the exception of $m_N$, we see that  all of the model parameters measured for both 
ROI's are in good agreement with each other.  This indicates that the shape of the excess 
does not change appreciably with the ROI and that our energy-dual scenario works well for 
both ROI's.  

The mismatch in the $m_N$ measurement is not surprising because only the upper 
and lower endpoints of the energy spectrum (\ie, the horizontal edges of the widest box) 
are sensitive to $m_N$, and it is precisely here where the 
signal statistics are relatively poor. An over- or under-estimate of 
the foreground/background flux could therefore easily shift both endpoints rather substantially.
Likewise, the best-fit values for $E_\ast$ for each ROI, while consistent with each other,
do not quite agree with the results of the
model-independent analysis in Sect.~\ref{sec:GCExcess}.~  However, as discussed above, the 
choice of energy-binning scheme may skew the model-independent results, as $E_\ast$ does not 
have to lie at the center of a given energy bin.  If this type of DDM scenario is realized 
in nature, one would expect the best-fit values for $E_\ast$ from the two analyses 
to more closely coincide as more data is acquired and as the energy resolution of the
detector is improved.

Finally, we observe that general features of our best-fit DDM models 
coincide nicely with the observed data.  As pointed out at the end of 
Sect.~\ref{sec:GCExcess}, the observed data is consistent with a relatively sharp peak in 
the photon spectrum near the global maximum --- a result which suggests that the 
injection spectrum $dN_\phi/dE_\phi$ of the intermediary particle $\phi$ remains non-zero as 
$E_{\phi} \rightarrow m_{\phi}$.  Furthermore, the rapidly falling nature of the photon 
energy spectrum, as shown in Fig.~\ref{fig:GC_flux}, suggests that the 
injection spectrum of $\phi$ should fall quickly as $E_{\phi}/ m_\phi$ increases.
In typical examples, the injection spectrum of $\phi$ particles produced from the 
annihilation/decay of a DDM ensemble typically follows a power law, as in 
Eq.~\eqref{eq:dNdEphi}.  One would therefore expect that the best fit to the Fermi-LAT 
data would arise from a falling power law ({\it i.e.}\/, from a negative scaling exponent 
$\xi < 0$).  The results obtained in our fit coincide with this expectation.

%%%%%%%%%%%%%%%%%%%%%%%%%%%%%%%%%%%%%%%%%%%%%%%%%%%%%%%%%%%%%%%%%%%%%%%%%%%%%%%

\section{Conclusions \label{sec:conclusion}}

%%%%%%%%%%%%%%%%%%%%%%%%%%%%%%%%%%%%%%%%%%%%%%%%%%%%%%%%%%%%%%%%%%%%%%%%%%%%%%%

The possibility that the excess in the flux of gamma rays emanating from the vicinity 
of the GC is the result of annihilating or decaying dark matter is an
intriguing one.  If dark matter is indeed responsible for this excess, 
one pressing question is what, if anything, we can learn about the properties
of the dark sector from the spectral information associated with that excess.
Dark-matter models of the gamma-ray excess typically rely on complicated cascade 
mechanisms for photon production in order to reproduce the spectrum of the excess ---
mechanisms whose non-trivial kinematics obscures the connection between the properties
of that spectrum and the properties of the dark-matter candidate.

In this paper, by contrast, we have considered an alternative dark-matter interpretation of the 
gamma-ray excess --- one in which a more direct connection exists
between the properties of the underlying dark sector
and the spectral shape of the gamma-ray excess to which it gives rise.
In particular, we have pointed out that the spectrum of the observed excess in the 
Fermi-LAT data is potentially invariant with respect to an energy duality transformation
of the form $E_\gamma \to E_\ast^2/E_\gamma$ for a self-dual 
energy $E_\ast \sim \mathcal{O}(1\mbox{~GeV})$.  Motivated by this observation, we
have presented a broad class of physical scenarios wherein such 
an energy self-duality is realized. 
In these scenarios, dark-matter annihilation/decay produces a non-trivial injection 
spectrum $dN_\phi/dE_\phi$ of intermediary particles $\phi$, each of which subsequently 
decays into a final state involving one or more photons which are mono-energetic and 
isotropically distributed in the $\phi$ rest frame.  We have also shown that an 
appropriate injection spectrum of $\phi$ particles for describing the Fermi-LAT data is 
naturally realized within the context of the DDM framework.  

It is clear that our scenario relies directly on the existence of a multi-component 
dark sector, as this is a primary ingredient of the DDM framework.
The possibility of non-minimal dark sectors has received increasing attention because 
many DM models predicated upon such sectors not only have non-trivial 
cosmological consequences (\eg, ``assisted freeze-out''~\cite{Belanger:2011ww}), but
also often interesting phenomenological implications as well  
(\eg, ``boosted dark matter''~\cite{Agashe:2014yua,Berger:2014sqa,Kong:2014mia}
as well as collider, direct-detection, and indirect-detection 
signatures~\cite{Dienes:2012yz, Dienes:2012cf, Dienes:2013xff, Dienes:2014bka, 
Dienes:2015bka} that transcend those normally associated with traditional WIMP-like 
single-component dark-matter scenarios).
Indeed, multi-component dark sectors can even give rise to enhanced complementarity 
relations which can be used to probe and constrain the parameter spaces of such 
models~\cite{Dienes:2014via}.
Thus, our explanation of the GC excess within the context of the DDM framework
--- if corroborated by future experiments --- could provide an interesting 
window into the physics of the dark sector.  Indeed,  it would be interesting to 
study the cosmological and phenomenological implications of the particular 
set of DDM parameters obtained in our fit to the Fermi-LAT data.

It is also important to realize that our discussion of the energy duality of the photon spectrum
under $E_\gamma \rightarrow E_\ast^2/E_\gamma$ has a broad applicability that extends
well beyond its application to the gamma-ray excess observed in the Fermi-LAT data.
Indeed, this duality can be used as a tool for deciphering the origins of {\it any}\/ 
generic continuum excess which might potentially be observed at future X-ray or 
gamma-ray facilities.  As discussed in Sect.~\ref{sec:Framework}, a broad
range of spectral shapes can be realized within scenarios of the sort described
above.  In particular, any bump-like feature in the gamma-ray spectrum can be realized 
in such a scenario, provided 
\begin{itemize}
  \item  the spectral feature is self-dual under the transformation 
    $E_\gamma \rightarrow E_\ast^2 / E_\gamma$;   and
  \item  the spectral feature has a global maximum at $E_\gamma = E_\ast$ and 
    decreases monotonically as $E_\gamma$ either increases or decreases away from 
    $E_\ast$.
\end{itemize}
Moreover, we have shown that in scenarios of this sort, the shape of the 
spectral feature is directly correlated with the behavior of the intermediary injection 
spectrum at $E_\phi = m_\phi$.  In particular, information about the kinematics of
$\phi$ production and decay is manifest in the behavior of $dN_\gamma/dE_\gamma$
near its maximum:  
\begin{itemize}
  \item If $dN_\phi / dE_\phi $ remains non-zero as 
    $E_\phi \rightarrow m_\phi$, the photon spectrum will  exhibit a cuspy peak at 
    $E_\ast$.
  \item If $dN_\phi / dE_\phi \rightarrow 0$ as 
    $E_\phi \rightarrow m_\phi$, the photon spectrum will be smooth at $E_\ast$.
\item If $dN_\phi / dE_\phi$ vanishes below some threshold energy
  $\overline{E} > m_\phi$, the photon spectrum will exhibit a plateau around 
  $E_\ast$.
\end{itemize}
Thus, an excess of photons emanating from any astrophysical source which possesses 
the above features not only lends itself to an interpretation in terms of our
annihilating/decaying dark-matter scenario, but can also yield additional information 
about the properties of the underlying dark sector.

In general, an intermediary particle $\phi$ which couples to photon pairs in the 
manner indicated in Eq.~(\ref{eq:CouplingLagrangian}) will also couple to gluon pairs 
through an interaction term of the form $\mathcal{L} \ni (c_g/f_{\phi}) \phi
G^a_{\mu\nu}\tilde{G}^{\mu\nu a}$, where $G^a_{\mu\nu}$ is the gluon field-strength 
tensor and $c_g$ is a 
dimensionless coefficient.  Such an interaction term leads to additional, hadronic 
decay channels for $\phi$.  The production of photons in association with these 
channels (through showering or from the decays of final-state hadrons) gives rise 
to an additional contribution to the differential photon flux.  The production rate 
for these photons depends on the value of $c_g$, which is highly model-dependent.  
In this paper, for simplicity, we have assumed that $c_g \ll 1$ and that this 
showering/hadron-decay contribution to $d\Phi/dE_\gamma$ is therefore negligible.  

It is nevertheless interesting to consider how our results would be modified in 
situations in which $c_g \sim \mathcal{O}(1)$ and the showering/hadron-decay 
contribution is significant.
In order to assess the impact of the showering/hadron-decay contribution on the 
overall photon signal spectrum from dark-matter annihilation in our DDM scenario, 
we begin by noting that this contribution, like the contribution from 
$\phi\rightarrow \gamma\gamma$ decay, owes its shape both to the kinematics of 
photon production in the rest frame of a decaying $\phi$ particle and to the 
spectrum of boosts imparted to the $\phi$ particles by the DDM ensemble constituents.  
The spectrum of boosts is governed in large part by the scaling exponent $\xi$, 
which characterizes how the contribution to the production rate of $\phi$ particles 
from an individual ensemble constituent $\chi_n$ scales with $m_n$ across the ensemble.  
As is evident from Fig.~\ref{fig:fit}, the best-fit values for $\xi$ for both of our 
ROI's are roughly $\xi \approx -2.5$, which implies that this contribution falls off 
rapidly with $m_n$.  As a result, we find that the {\it collective}\/ contribution to 
the production rate for secondary photons from ensemble constituents $\chi_n$ with 
$m_n$ above a few GeV is essentially negligible, even when $c_g \sim \mathcal{O}(1)$.  
Thus, the contribution to the overall signal flux from showering/hadron decay is 
expected to be significant only for $E_\gamma \lesssim \mathcal{O}(1\mathrm{~GeV})$.    
    
The kinematics of photon production from showering/hadron-decay in the rest frame 
of the decaying intermediary has important ramifications as well.  The primary parameter 
of interest here is $m_\phi$, our best-fit value for which is $m_\phi \approx 1.2$~GeV 
for both ROI's.  Since this is less than twice the proton mass, baryon-number 
conservation implies that final states consisting primarily of light mesons --- and 
especially of pions --- should dominate the partial width of $\phi$ to hadrons.  
The dominant contribution to the secondary-photon spectrum at 
$E_\gamma \sim \mathcal{O}(1\mathrm{~GeV})$ is therefore likely to be the 
contribution from on-shell $\pi^0\rightarrow \gamma\gamma$ decay.  Photons produced in 
this way have their own distinctive kinematics.  In particular, the energy spectrum 
associated with these photons manifests an energy duality of its own, with 
self-dual energy $m_{\pi^0}/2$.  The presence of such a duality could be exploited 
in order to disentangle this contribution from the primary photon spectrum.  
In principle, one could significantly reduce the contamination of the primary 
spectrum by subtracting off the contribution to the signal flux which is dual under 
$E_\gamma \rightarrow \m_{\pi^0}^2/(4E_\gamma)$.  Moreover, since the shape of this 
$\pi^0$-decay contribution to the secondary photon spectrum is correlated with the 
shape of the primary-photon spectrum, a comparison between these two contributions 
could provide additional evidence in support of a DDM origin for the GC excess.
Indeed, this strategy has been successfully employed within the context of other, 
similar DDM scenarios~\cite{MeVDDM}.   
Such an analysis would of course require improved data on the gamma-ray spectrum 
at energies $E_\gamma \lesssim m_{\pi^0}/2$.  However, several proposals for 
instruments which would provide significant improvements in energy resolution 
within that energy range have been advanced~\cite{Boggs:2006mh,ASTROGAM}.  Thus, we see 
that ``contamination'' from the showering/hadron-decay contribution to the differential 
photon flux that arises in this DDM scenario when $c_g \sim \mathcal{O}(1)$ may 
actually be an asset rather than a hurdle in the effort to distinguish this scenario 
from other models for the origin of this excess.

One final comment is in order.  In particular,
we stress that although the gamma-ray excess observed in the Fermi-LAT data is 
{\it consistent}\/ with an energy duality of the kind we have discussed in this paper, 
there are significant uncertainties in the spectral shape of the excess which, at
present, preclude any more definitive statements along these lines.  These  
include not only statistical uncertainties, but also systematic uncertainties in the 
astrophysical foregrounds/backgrounds in the vicinity of the GC and uncertainties 
resulting from the energy resolution of the the Fermi-LAT instrument.  Moreover, 
the preferred value for the self-dual energy $E_\ast \sim \mathcal{O}(1\mathrm{~GeV})$ 
is very close to the lower limit of the energy range for which reliable data exists.   
As a result, current data does not yet permit us to distinguish between
the annihilating/decaying dark-matter scenario we have described here and other 
possible explanations of the CG gamma-ray excess.  However, there are new astronomical 
instruments, both planned and under consideration, which are far better equipped to
investigate whether the gamma-ray spectrum from the GC indeed exhibits such an 
energy-duality.  For example, GAMMA-400 is expected to have a better energy resolution 
than Fermi-LAT in the $E_\gamma \sim 1~\gev$ regime.  A variety of instruments
designed 
to study the gamma-ray spectrum in the $10~\mev\lesssim E_\gamma \lesssim 1~\gev$ regime, 
such as ASTROGAM~\cite{ASTROGAM}, 
have also recently been proposed, often with energy resolutions far superior to those of
similar experiments past or present.  High-statistics data from such experiments 
could potentially definitively rule out or else lend significant credence to our scenario.
Indeed, this illustrates that even when an excess of photons observed at indirect-detection
experiments has the form of a broad continuum bump, precision measurements of the spectral 
shape of this bump can prove crucial for our understanding of the underlying physics.

\bigskip

%%%%%%%%%%%%%%%%%%%%%%%%%%%%%%%%%%%%%%%%%%%%%%%%%%%%%%%%%%%%%%%%%%%%%%%%%%%%%%%

\section*{Acknowledgments}

%%%%%%%%%%%%%%%%%%%%%%%%%%%%%%%%%%%%%%%%%%%%%%%%%%%%%%%%%%%%%%%%%%%%%%%%%%%%%%%

\vspace{-0.2cm}
We are grateful to Regina Caputo and Jeff Kost for useful discussions.
We would also like to acknowledge the Center for Theoretical Underground Physics 
and Related Areas (CETUP$^\ast$) for hospitality and partial support during the 2016 
Summer Program.  KB and JK are supported in part by the National Science Foundation 
under CAREER Grant PHY-1250573, while KRD is supported in part by the Department of 
Energy under Grant DE-FG02-13ER41976 and by the National Science Foundation through 
its employee IR/D program.  DK was supported in part by the U.S.\ Department of Energy under 
Grant DE-SC0010296 and is presently supported 
in part by the Korean Research Foundation (KRF) through the CERN-Korea 
Fellowship program. 
JCP is supported by the National Research Foundation of Korea
(NRF-2013R1A1A2061561, 2016R1C1B2015225).  The opinions and conclusions expressed 
herein are those of the authors, and do not represent any funding agencies.

\appendix
\section{~~Calculating the Photon Flux of the DDM Model}

In this Appendix we provide
a rigorous calculation of the differential photon flux $d\Phi/dE_\gamma$
corresponding to the DDM model introduced in Sect.~\ref{modelsect}.
This derivation will also confirm the normalization factors introduced in Eq.~(\ref{templatefct})
and likewise clarify the meaning of the ``effective'' differential 
number $dN_\phi/dE_\phi$
given in Eq.~(\ref{eq:dNdEphi}).
By and large, our approach will generally follow that of Ref.~\cite{MeVDDM}.

We begin by noting that while the expression for the differential flux
$d\Phi/dE_\gamma$ in Eq.~(\ref{Jversion}) is suitable for a single dark-matter
candidate $\chi$, in multi-component contexts involving fields $\chi_n$ with $n=0,1,...,N$
this expression generalizes to take the form
\beq
     {d\Phi_n\over dE_\gamma} ~=~ {J\over 4\pi} \, {\langle \sigma v \rangle_n \over 4 m_n^2}\, 
                 {\Omega_n\over \Omega_{\rm tot}} \, {d N^{(n)}_\gamma\over dE_\gamma}~.
\label{app1}
\eeq
Here $\Phi_n$ and $\Omega_n$ are respectively the flux contribution and cosmological abundance of $\chi_n$, with
$\Phi=\sum_{n=0}^N \Phi_n$ and $\Omega_{\rm tot}\equiv  \sum_{n=0}^N \Omega_n$.
Likewise we observe that 
\beq
     \Phi_0 ~=~ {J\over 4\pi} \, {\langle \sigma v \rangle_0 \over 4 m_0^2}\, 
                 {\Omega_0\over \Omega_{\rm tot}} \, N_\gamma^{(0)}~
\eeq
where $N_\gamma^{(0)}=4$
is the number of photons produced through the annihilation of $\chi_0$.
We can therefore write Eq.~(\ref{app1})
in the form
\beq
     {d\Phi_n\over dE_\gamma} ~=~ {\Phi_0\over 4} \,
     \, \left( {m_0^2\over m_n^2} \, {\Omega_n\over \Omega_0} 
          \, {\langle \sigma v\rangle_n\over  \langle \sigma v\rangle_0}\right)
       \, {dN_\gamma^{(n)}\over dE_\gamma}~.
\label{app2}
\eeq
Given this, the DDM scaling behavior in Eq.~(\ref{eq:scalingbeh})
implies that the quantities within the parentheses in Eq.~(\ref{app2})
each scale in such a way that this entire parenthesized factor
is equal to $(\sqrt{s_n}/\sqrt{s_0})^\xi$.
Summing over the modes of the ensemble 
and passing to the continuous integral form of the sum
as described above Eq.~(\ref{eq:scalingbehcont})
then yields
\beq
    {d\Phi\over dE_\gamma} ~=~ {\Phi_0\over 4\Delta(\sqrt{s})} \int_{\sqrt{s_0}}^{\sqrt{s_N}} d\sqrt{s}  
          \left( {\sqrt{s}\over \sqrt{s_0}}\right)^\xi 
         {dN_\gamma\over dE_\gamma}~.
\label{app3}
\eeq
However, the differential photon number $dN_\gamma/dE_\gamma$ is given in Eq.~(\ref{eq:GenSpectrum}).
Substituting this expression into Eq.~(\ref{app3}) and recognizing that $E_\ast= m_\phi/2$, we thus have
\beqn
    {d\Phi\over dE_\gamma} &=&  {\Phi_0\over 2 \Delta(\sqrt{s})} \int_{\sqrt{s_0}}^{\sqrt{s_N}} d\sqrt{s}  
          \left( {\sqrt{s}\over \sqrt{s_0}}\right)^\xi \nonumber\\
  && ~\times  
   \int_{\frac{m_\phi}{2} (x+1/x)}^\infty dE_\phi
          \frac{1}{\sqrt{E_\phi^2 - m_\phi^2}}
    \, \frac{dN_\phi}{dE_\phi}  ~.~~~~~~
\label{app4}
\eeqn
 
In Eq.~(\ref{app4}), the final quantity is the differential number 
$dN_\phi/dE_\phi$.
Strictly speaking, this is given by
\beq
               {dN_\phi\over dE_\phi} ~=~ 2 \, \delta (E_\phi - \sqrt{s}/2)~,
\label{Diracdelta}
\eeq
signifying that the annihilation of each dark-sector component
with CM energy $\sqrt{s}$ 
produces exactly two $\phi$ particles with energies $E_\phi=\sqrt{s}/2$.
However, it is possible to rewrite Eq.~(\ref{app4}) by exchanging the order of integrations,
yielding
\beqn
    {d\Phi\over dE_\gamma} &=&  {\Phi_0\over 2 \Delta(\sqrt{s})} 
   \int_{\frac{m_\phi}{2} (x+1/x)}^\infty dE_\phi
          \frac{1}{\sqrt{E_\phi^2 - m_\phi^2}} \nonumber\\
 && \times \int_{\sqrt{s_0}}^{\sqrt{s_N}} d\sqrt{s}  
          \left( {\sqrt{s}\over \sqrt{s_0}}\right)^\xi 
           \frac{dN_\phi}{dE_\phi}~.  
\label{app5}
\eeqn
Given this, we may identify the entire quantity on the second line of Eq.~(\ref{app5})
as an ``effective'' $dN_\phi/dE_\phi$, one which combines not only the 
Dirac $\delta$-function contribution in Eq.~(\ref{Diracdelta}) but also the
scaling factor $(\sqrt{s}/\sqrt{s_0})^\xi$.  
Indeed, this is precisely the quantity which was
constructed in Eq.~(\ref{eq:dNdEphi})
and which (by abuse of notation) was casually denoted $dN_\phi/dE_\phi$ throughout the body
of the text.
As such, it is this effective quantity which encodes not only the widths but also the heights
of the stacked ``boxes'' in Fig.~\ref{fig:stacking}.

Eq.~(\ref{app2}) also affords us another way of interpreting this effective
number $dN_\phi/dE_\phi$.
In Eq.~(\ref{app2}), it is the quantity in parentheses which varies across the
ensemble and which, in so doing, exhibits the DDM scaling behavior.
However, for the purposes of calculating fluxes, we can equivalently imagine that the quantities 
within the parentheses in Eq.~(\ref{app2}) are actually constant, and that their  scaling 
behavior has been absorbed into an effective differential number $dN_\phi/dE_\phi$ instead.
Indeed, it is precisely for these reasons that 
a simple relation such as that in Eq.~(\ref{ratios}) holds when
written in terms of effective number densities.

Given the expression in Eq.~(\ref{app5}), evaluation of 
the flux $d\Phi/dE_\gamma$ now proceeds directly.
This then yields the results listed in Sect.~\ref{modelsect}
and confirms the overall normalizations 
quoted there.

%%%%%%%%%%%%%%%%%%%%%%%%%%%%%%%%%%%%%%%%%%%%%%%%%%%%%%%%%%%%%%%%%%%%%%%%%%%%%%%

\end{document}